\documentclass[useAMS,usenatbib]{mn2e}
\usepackage[utf8]{inputenc}

\usepackage{threeparttable}
\usepackage{graphicx, color}
\usepackage[svgnames]{xcolor}
\usepackage[]{txfonts}
\usepackage{hyperref}
\usepackage{soul}

\newcommand{\eqb}{\begin{eqnarray}}
\newcommand{\eqe}{\end{eqnarray}}

\newcommand{\mpr}{m_{\rm p}}

\usepackage{epsfig}
\usepackage{epstopdf}
\usepackage{bm}

\title[Wind - Disk Interactions: Sgr~A*]{Modelling Accretion Disk and Stellar Wind Interactions: the Case of Sgr~A$^\star$}
\author[Christie et al.]{I.~M. Christie$^{1}$\thanks{E-mail:
ichristi@purdue.edu}, M.~Petropoulou$^1$\thanks{Einstein Fellow}, P.~Mimica$^{2}$\thanks{Email: petar.mimica@uv.es}, D.~Giannios$^{1}$\\
$^{1}$Department of Physics and Astronomy, Purdue University, 525 Northwestern
Avenue, West Lafayette, IN 47907, USA \\
$^{2}$Departament d'Astronomia i Astrof\'{\i}sica
Universitat de Val\`{e}ncia}

\begin{document}

\maketitle
\begin{abstract}
Sgr A* is an ideal target to study low-luminosity accreting systems. It has been recently proposed that properties of the accretion flow around Sgr A* can be probed through its interactions with the stellar wind of nearby massive stars belonging to the S-cluster. When a star intercepts the accretion disk, the ram and thermal pressures of the disk terminate the stellar wind leading to the formation of a bow shock structure. Here, a semi-analytical model is constructed which describes the geometry of the termination shock formed in the wind. With the employment of numerical hydrodynamic simulations, this model is both verified and extended to a region prone to Kelvin-Helmholtz instabilities. Because the characteristic wind and stellar velocities are in $\sim10^{8}$ cm s$^{-1}$ range, the shocked wind may produce detectable X-rays via thermal bremsstrahlung 
emission. The application of this model to the pericenter passage of S2, the brightest member of the S-cluster, shows that the shocked wind produces roughly a month long X-ray flare with a peak luminosity of $L\approx 4 \times 10^{33}$ erg s$^{-1}$ for a stellar mass-loss rate, disk number density, and thermal pressure strength of $\dot{M}_{\rm w}= 10^{-7} M_\odot\, {\rm yr}^{-1}$, $n_{\rm d} = 10^{5}$ cm$^{-3}$, and $\alpha=0.1$, respectively. This peak luminosity is comparable to the quiescent X-ray emission detected from Sgr A* and is within the detection capabilities of current X-ray observatories. Its detection could constrain the density and thickness of the disk at a distance of $\sim 3000$ gravitational radii from the supermassive black hole.
\end{abstract}


\section{Introduction}
\label{intro}
Stellar wind bow shocks form due to the supersonic transit of outflowing gas from a star while interacting with an ambient medium. The formation of such structures has been studied extensively, ranging from colliding winds in a binary system \citep{stevens1992, canto1996}, to interactions between the stellar wind of a single star with an ambient medium \citep[ henceforth W96]{canto1996, wilkin1996}, to Modelling heliospheric structures around the heliopause
\citep{drake2015, washimi2015}, as well as non-thermal particle acceleration at the shock front \citep{guo2014}. Studying the post-shock emission provides a means to probe different properties of both the stellar wind and the ambient medium (e.g. density, velocity, temperature).

The compact radio source, Sgr A*, marks the location of the supermassive black hole at our Galactic center. Accretion onto the black hole is believed to power the radio, IR, and X-ray emission produced around Sgr A* \citep{genzel2010}. Located near the Galactic center is a group of early, type-B stars known as the S-cluster. These stars have been the target of extensive observational effort in an attempt to identify their evolutionary stage and origin \citep{eisenhauer2005, ghez2005, gillessen2009a}. The orbits of many S-stars are likely to intercept and interact with the accretion disk surrounding Sgr~A*. The brightest member of the S-cluster, S2, is a massive, main-sequence star with a mass-loss rate of $\dot{M}_{\rm w} \leq 3 \times 10^{-7} M_\odot\, {\rm yr}^{-1}$ \citep{martins2008} and is characterized  by its tight orbit with a pericenter distance of $\sim 3000$ $R_{\rm g}$, where $R_{\rm g}$ is the gravitational radius \citep{gillessen2009b}. Interactions between the disk gas and the stellar wind 
will 
cause a bow-shock structure to form, eventually engulfing the star. Given that the characteristic velocities of the wind and disk fluid are a few $10^{8}$ cm s$^{-1}$, the shocked fluid is expected to radiate in the X-ray band.

The use of S2 as a probe of accretion disk properties (e.g density and temperature) of Sgr A* was first proposed by \citealt[][]{giannios2013}, hereafter GS13. By comparing the ratio of the expansion timescale to the bremsstrahlung cooling timescale of the shocked stellar wind while adopting a spherically symmetric model, they were able to estimate the radiated, thermal, X-ray luminosity. The peak emission was predicted to take place around the pericenter passage of the star, where the ram pressure from the disk becomes maximal, with the X-ray luminosity potentially exceeding the quiescent emission from Sgr A*. A more detailed calculation of the radiated power is required as the next pericenter passage of S2 is expected in 2018.

In this study, an expansion is made upon the previous work of GS13 by constructing a two-dimensional, semi-analytical model for the shocked stellar wind region. To do so, axisymmetry is assumed while following a similar analysis of momentum supported bow shocks presented by W96 which emphasizes the conserved momentum flux within the shocked region. With the inclusion of thermal pressure from the disk, the shape of the shock surface is then derived as a function of $\alpha$, i.e. the ratio of the thermal pressure to the ram pressure of the disk. Here, the thermal pressure of the disk is expected to be non-negligible due to the disk being radiatively inefficient and partially pressure supported. Quantities downstream from the termination shock (e.g. density, temperature, pressure) may then be determined by using the Rankine-Hugoniot conditions. The semi-analytic formalism adopted for the description of the system allows for a more accurate computation of the thermal bremsstrahlung radiated power produced by 
the shocked stellar wind.

Since the semi-analytical model is built upon a series of assumptions, the validity of these results is tested through detailed hydrodynamic simulations. The presented model is found to provide a good description of the shock surface and of the thermodynamical properties in the downstream region of the forward shock up to an angle $\theta\approx \pi/2$, where $\theta$ is measured with respect to the symmetry axis. At larger angles and at larger distances in the tail of the bow shock, the semi-analytical model is not able to describe the mixing of the shocked disk and stellar wind fluids caused by Kelvin-Helmholtz instabilities. The role of the latter on the radiative output of the mixed fluids is, therefore, investigated solely through numerical hydrodynamic simulations. These simulations are also used to investigate, more realistically, the time-dependent problem: namely a situation where the disk density changes on a timescale similar to the time it takes for the star to complete its pericenter passage.

This paper is structured as follows. In Sec. \ref{model}, the mathematical formulation of the system of colliding fluids is presented along with the semi-analytical results. In Sec. \ref{numerical}, a discussion is given on the employment of numerical hydrodynamic simulations used in order to test the validity of the model as well as to determine the thermal radiated power produced by the shocked wind regions. In Sec. \ref{sec:application}, this model is applied to the accretion disk of Sgr A* and specific S-stars while putting an emphasis on S2. By Modelling the pericenter transit of S2 as it precesses through the accretion disk of Sgr A*, we will make predictions of possible, observational signatures. A discussion and summary of the results are then given in Sec. \ref{sec:discussion} and Sec. \ref{sec:summary} respectively.

\section{The model}
\label{model}
In order to derive the shape of the shocked stellar wind region, it is assumed that the wind-disk system has reached a steady-state\footnote{The steady state assumption is relaxed in Sec.~\ref{sec:application} where time-dependent disk properties are simulated numerically.} and that the shocked wind region can be described as a thin shell (i.e., $H<<R$ where $H$ and $R$ are the shell's width and radial distance from the star). Within the thin-shell approximation, the contact discontinuity, which separates the shocked stellar wind from the shocked disk material, and the termination shock of the stellar wind coincide.

In the rest frame of the star, the stellar wind is assumed to be isotropic with a mass-loss rate of $\dot{M}_{\rm w}$ and constant velocity $\bm{v}_{\rm w}=\bm{\hat{e}}_{\rm r} v_{\rm w}$, with the unit vector denoting the spherical radial direction. The disk fluid has a constant mass density $\rho_{\rm d}$ and, in the rest frame of the star, moves with a speed $\bm{v}_{\rm d}=-\bm{V}_\star$, where $V_\star$ is the velocity of the star.
Let the $z$-axis be the symmetry axis of the system and the motion of the star as it precesses through the disk to be in
the $\bm{\hat{e}}_{\rm z}$ direction. In the rest frame of the star, the gas from the disk approaches the star from the $-\bm{\hat{e}}_{\rm z}$ direction.
The standoff distance of the termination shock
is determined by balancing the pressures from both sides of the shocked region. The ram pressure of the stellar wind decreases with increasing radius $R$ as $P_{\rm w} = \dot{M}_{\rm w} v_{\rm w}/4 \pi R^2$. The disk pressure is the sum of the thermal and ram pressures, the latter being a result of the relative motion of the disk and the star. The ram pressure of the disk is $P_{\rm d} = \rho_{\rm d} v_{\rm d}^{2}$ while the thermal pressure is defined as a multiple of the ram pressure, $P_{\rm th} = \alpha P_{\rm d}$. Values of $\alpha$ within the range $0 \le \alpha \leq 1$ are considered\footnote{For $\alpha \ge 0.6$, a shock does not form in the disk, since the motion of the star through the hot fluid of the disk becomes subsonic; if $\gamma=5/3$ is the adiabatic index of the disk fluid, then $c_{\rm s}/v_{\rm d} = \sqrt{\gamma \alpha} > 1$ for $\alpha>1/\gamma$. The stellar wind, however, is always terminated through a strong shock.}, with $\alpha=0$ representing a cold disk as described by W96. By 
matching the pressures, the standoff distance is
\eqb
R_{0} = \sqrt{\frac{\dot{M}_{\rm w} v_{\rm w}}{4 \pi (1 + \alpha) \rho_{\rm d} v_{\rm d}^{2}}},
\label{Ro}
\eqe
which sets the characteristic length scale of the system. The distance $R_1$ of the termination shock in the opposite direction of the standoff point is
determined by balancing the ram pressure from the unshocked wind with the pressure of the shocked wind, the latter estimated to be the thermal pressure of the disk. This may be written in terms of $R_0$ as $R_1 = R_0 \sqrt{1+\alpha^{-1}}$.

The mass and total momentum flux traversing an annulus of the shocked region are $2\pi$ times the following
\eqb
\label{Phim}
\Phi_{\rm m} & = &  R \sigma v_{\rm t} \sin\theta, \\
\Phi_{\rm t} & = &  R \sigma v_{\rm t}^{2} \sin\theta
\label{Phit}
\eqe
where $\sigma$ and $v_{\rm t}$ are the mass surface density and the tangential speed of the fluid within the shocked region, respectively.
In a steady-state situation, the mass flux moving within a cross sectional ring of the shell is equal
to the sum of the mass flux imparted from the wind on the solid angle occupied by the shocked region of the forward shell plus the mass flux from the
disk hitting the circular area of the projected cross section of the shell
\eqb
\label{Phim2}
\Phi_{\rm m} = \frac{\dot{M}_{\rm w} (1 - \cos\theta)}{4 \pi} + \frac{1}{2} R^{2} \rho_{\rm d} v_{\rm d} \sin^{2}\theta.
\eqe

The flow of shocked gases within the shell depends on the incoming momentum fluxes.
Let us consider a small wedge of constant width  $\Delta \phi$  in the azimuthal direction about the symmetry axis.
The rate at which vector momentum is being imparted on the wedge of the shell by the stellar wind is
\eqb
\bm{\Phi}_{\rm w}(\theta) \Delta \phi &=&\Delta \phi\int_{0}^{\theta} \rho_{\rm w}  R^2 \bm{v}_{\rm w}\bm{v}_{\rm w}\cdot\bm{\hat{e}}_{\rm r} \sin\theta' {\rm d}\theta'= \nonumber \\
&=&  \Delta \phi \frac{\dot{M}_{\rm w} v_{\rm w}}{8 \pi} \left[(\theta -\sin\theta \cos\theta) \ \bm{\hat{e}}_{\varpi} + \sin^{2}\theta \ \bm{\hat{e}}_{\rm z}\right]
\label{wind_momentum}
\eqe
where $\rho_{\rm w} = \dot{M}_{\rm w}/ (4 \pi R^2 v_{\rm w})$ was used and $\varpi=R\sin\theta$ denotes the cylindrical radial coordinate. The integration was performed on a spherical surface with unit element ${\rm d}\bm{A} = \bm {\hat{e}}_{\rm r}R^2\sin\theta {\rm d}\theta$, which is valid as long as the surface integral of the wedge region is independent of the shell's shape, $R(\theta)$. The latter is true thanks
to the fact that the system is momentum conserving, as first mentioned in W96.
\begin{figure*}
\begin{center}
\includegraphics[height=0.33\textwidth]{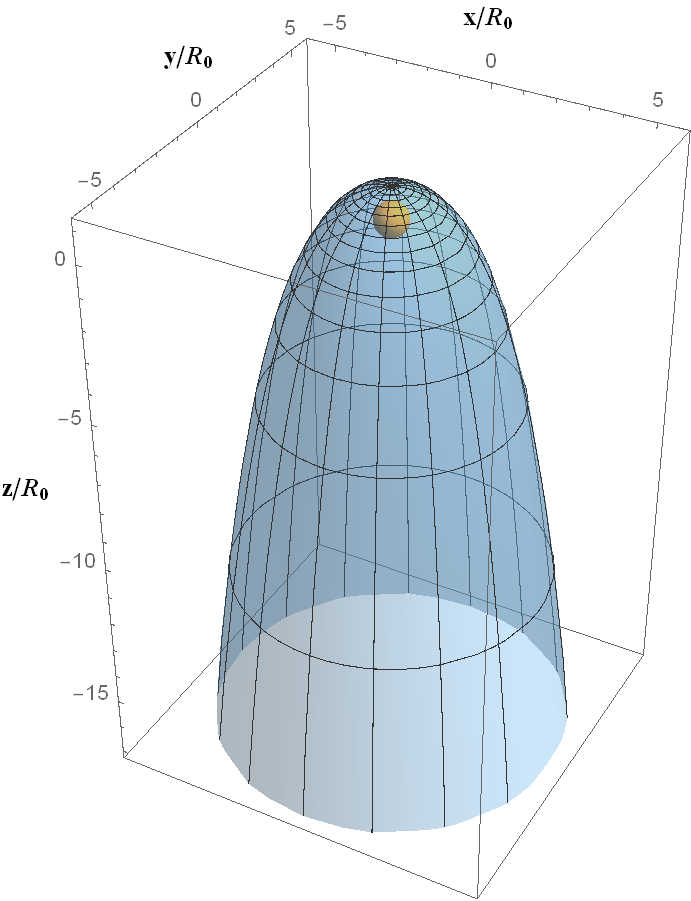}
\includegraphics[height=0.33\textwidth]{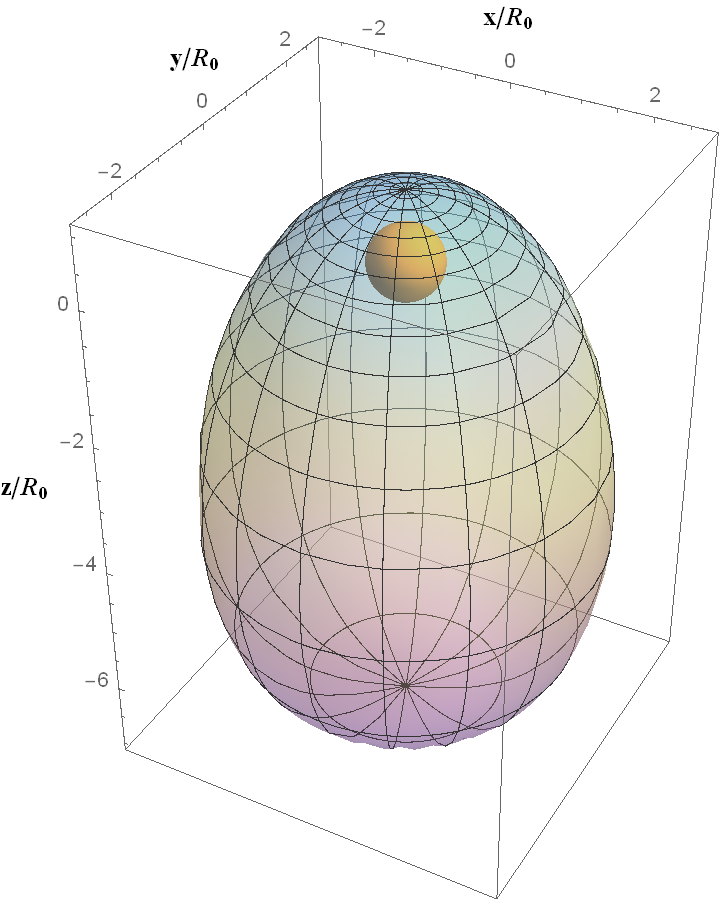}
\includegraphics[height=0.33\textwidth]{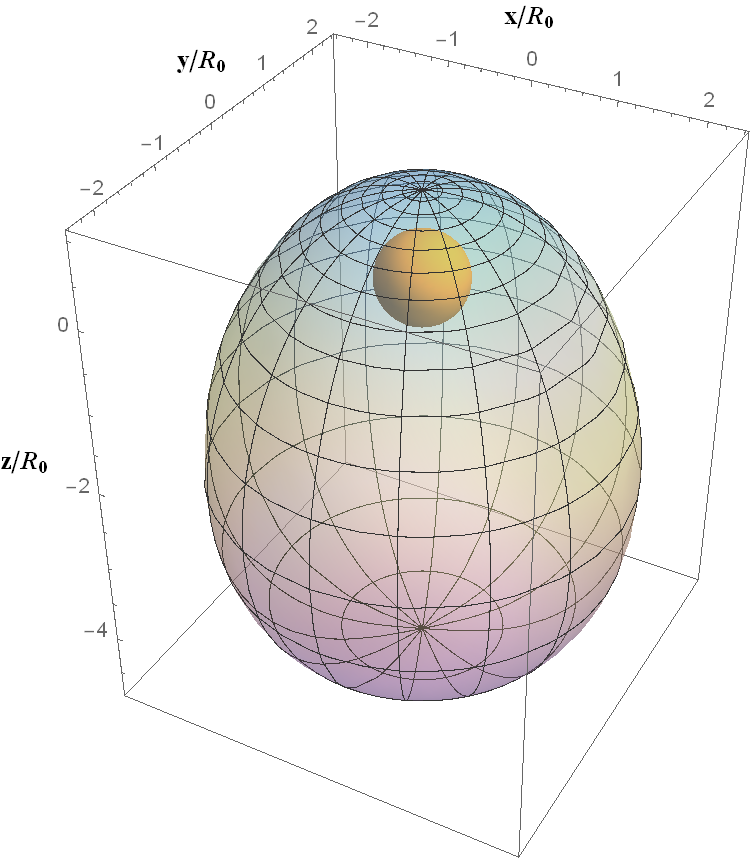}
 \caption{A three-dimensional view of the termination shock surface. From left to right $\alpha=10^{-5},0.05$, and $0.1$. A sphere of fixed size, centered at the position of the star, is overplotted in order to demonstrate the different scales among the structures. }
 \label{fig:surfaces}
\end{center}
\end{figure*}

The rate at which vector momentum is being imparted on the same wedge by the accretion disk is
\eqb
\bm{\Phi}_{\rm d}(\theta)\Delta \phi= \Delta \phi\int_{0}^{\theta} \left[-P_{\rm th} \bm{I} + \rho_{\rm d} \bm{v}_{\rm d}\bm{v}_{\rm d}\right]\cdot \bm{\hat{e}}_{r}R^{2} \sin\theta' {\rm d}\theta'= \nonumber \\  -\Delta \phi\frac{R^{2} \rho_{\rm d} v_{\rm d}^{2}}{2} \left[ \alpha (\theta -\sin\theta \cos\theta) \ \bm{\hat{e}}_{\varpi} + (1+ \alpha) \sin^{2}\theta \ \bm{\hat{e}}_{\rm z}\right]
\label{disk_momentum}
\eqe
where $\bm{I}$ is the identity tensor. If the thermal pressure of the disk is neglected, i.e. $\alpha =0$, the result of W96 is recovered.

The rate at which the total vector momentum is being deposited is therefore the sum of the individual contributions, namely
\eqb
\bm{\Phi}_{\rm t}(\theta) = \frac{\dot{M}_{\rm w} v_{\rm w}}{8 \pi} \left[ \bm{\hat{e}}_{\varpi} (\theta - \sin\theta \cos\theta) (1 - \lambda r^{2}) + \bm{\hat{e}}_{z} (1 - r^2) \sin^{2}\theta \right]
\label{total_momentum}
\eqe
where $\lambda$ is defined as
\eqb
\lambda \equiv \frac{\alpha}{1 + \alpha}
\label{lambda}
\eqe
and $r$ is the dimensionless radial distance defined as
\eqb
r(\theta) \equiv \frac{R(\theta)}{R_{0}}.
\label{r}
\eqe
The inclusion of the thermal pressure in the disk results in a decrease of the cylindrical radial component of the total momentum flux imparted on the shell. This affects the shape of the shell in such a way that, if $\alpha$ is non-negligible, the termination shock will enclose the star and will no longer extend out to infinity, as found by W96 for the case of a cold disk.

\subsection{The shape of the termination shock}
The key point stressed by W96 is that the shape of the shell is not determined by the local balance of different pressures acting on it, but by the condition that the total vector momentum flows in the direction tangential to the shocked region (i.e. the natural direction).
To determine the shape of the termination shock as a function of $\theta$, a transformation of coordinates is made from $(\varpi, z)$ to $(r, \theta)$ by using
${\rm d}z/{\rm d}\varpi=\Phi_{\rm t,z}/\,\Phi_{\rm t,\varpi}$. This leads to
\eqb
\frac{r^\prime \cos\theta-r\sin\theta}{r^\prime \sin\theta + r \cos\theta} = \frac{(1 - r^2) \sin^{2}\theta}{ (\theta - \sin\theta \cos\theta) (1 - \lambda r^{2})},
\label{transformation}
\eqe
where $r^\prime \equiv {\rm d}r /{\rm d}\theta$. A rearrangement of eq.~(\ref{transformation}) yields the following differential equation for $r(\theta)$
\eqb
r^\prime= \frac{r \sin\theta \left[ \theta (1 - \lambda r^{2}) + r^{2} (\lambda - 1) \sin\theta \cos\theta \right]}{(1 - \lambda r^{2}) (\theta - \sin\theta \cos\theta) \cos\theta - \sin^{3}\theta (1 - r^2)}
\label{ode}
\eqe
which can be solved numerically for different values of $\lambda$ with  $r(0)$=1 as an initial condition.  By setting $\lambda=0$, the solution of the equation above reduces to $r(\theta) = \csc\theta\sqrt{3(1-\theta\cot\theta)}$, as given by W96.

\begin{figure}
 \centering
\includegraphics[height=0.4\textwidth]{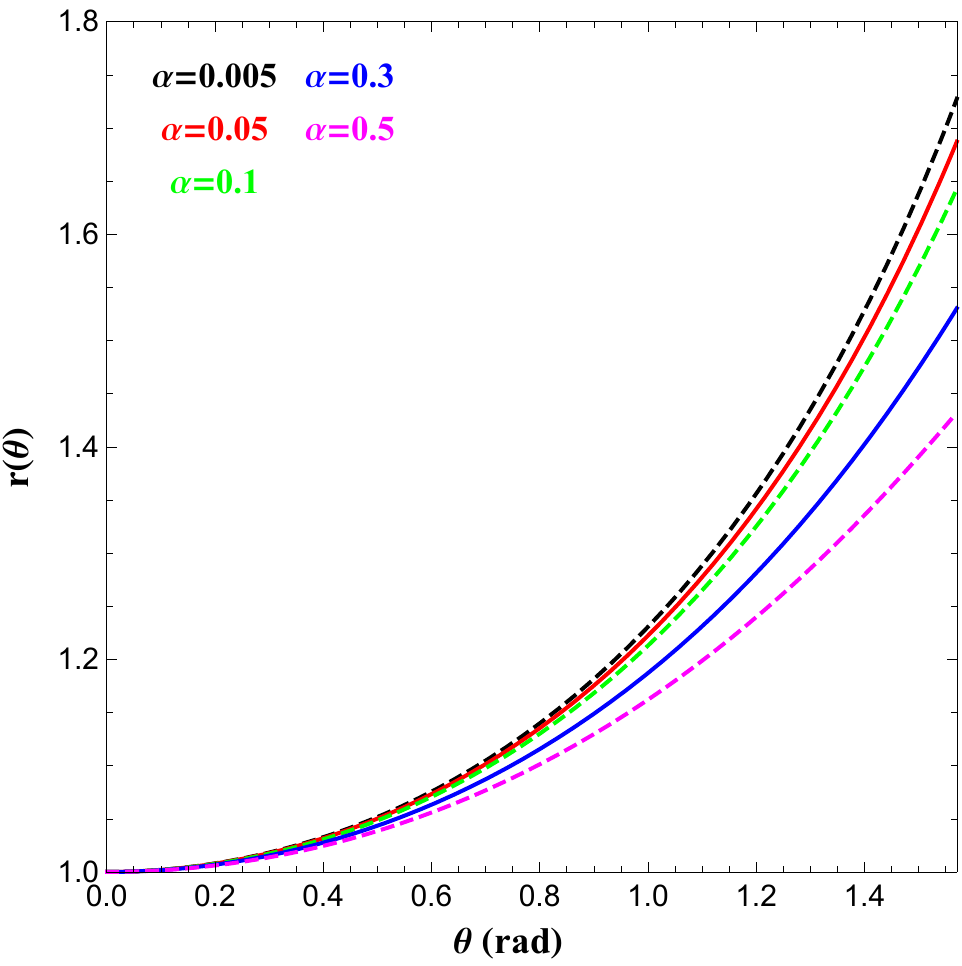}
\caption{Plot of the termination shock distance $R$, normalized to $R_{0}$, as a function of $\theta$ for the $\alpha$ values marked on the plot. These values correspond to $\lambda$ values of $0.005$, $0.05$, $0.1$, $0.23$, and $0.35$.}
\label{fig:r}
\end{figure}

Figure \ref{fig:surfaces} shows a three-dimensional view of the termination shock surface obtained after solving eq.~(\ref{ode}) for
$\alpha=10^{-5}$, $0.05$ and $0.1$, under the assumption of axisymmetry in the $\bm{\hat{e}_\phi}$ direction. To illustrate the different
scales among the three cases, a sphere of fixed size is overplotted and centered at the position of the star. The radius of the sphere does not represent the actual radius of the star.  The increasing thermal pressure of the surrounding medium drives $R_0$ closer towards the star while confining the shock into a closed, oval-shaped surface. Figure \ref{fig:r} shows the radial dependence of the termination shock obtained by solving eq.~(\ref{ode}) for various $\alpha$ values up to an angle of $\pi/2$. The reasons for choosing the particular $\theta$ range will become apparent in Sec.~\ref{numerical}.
The distance of the shock increases already by $10\%$ with respect to the standoff value at $\theta \simeq \pi/5 - \pi/4$, for all $\alpha$ values.
However, only  for $\alpha > 0.11$ do the effects of the disk's thermal pressure on the shape of the shock surface at $\theta=\pi/2$ become evident.

\subsection{The shocked stellar wind}
\label{ssw}
The properties of the shocked, stellar wind, e.g. temperature and number density, are crucial for the calculation of the radiated thermal bremsstrahlung power. These properties are determined by using the Rankine-Hugoniot jump conditions across the shock front with the assumption that the shape of the termination shock is adequately described by $r(\theta)$ and that the stellar wind is an ideal gas with a specific heat ratio  $\gamma=5/3$.

The application of the jump conditions requires the knowledge of the normal and tangential unit vectors along the shock surface. The latter is obtained by knowing that $\bm{\Phi}_{\rm t}$ flows tangential to surface
\eqb
\label{unit_tangent}
\bm{\hat{n}}_{\rm t} \equiv \frac{\Phi_{\rm t}}{|\Phi_{\rm t}|}= A_\varpi(\theta) \bm{\hat{e}}_{\varpi} + A_{\rm z}(\theta) \bm{\hat{e}}_{\rm z},
\eqe
where
\eqb
\label{Api}
 A_\varpi(\theta) & = & \frac{(\theta-\sin\theta\cos\theta)(1-\lambda r^2)}{f_\lambda(\theta)} \\
 A_{\rm z}(\theta) & = & \frac{\sin^2\theta(1-r^2)}{f_\lambda(\theta)},
 \label{Az}
\eqe
and $f_{\lambda}$ is defined as
\eqb
\label{f}
f_\lambda(\theta) = \left[(\theta - \sin\theta \cos\theta)^{2} (1 - \lambda r^{2})^{2} + \sin^{4}\theta (1 - r^{2})^{2}\right]^{1/2}.
\eqe
The normal unit vector outward to the termination shock is then determined from eq.~(\ref{unit_tangent}) and is written as
\eqb
\bm{\hat{n}}_{\bot}= -A_{\rm z}(\theta) \bm{\hat{e}}_\varpi + A_\varpi(\theta)\bm{\hat{e}}_{\rm z}.
\eqe
All quantities of the shocked stellar wind, i.e. in  the downstream of the shock,  will be noted with the subscript `sw'.
The mass density $\rho_{\rm sw}$ of the shocked stellar wind is given by
\eqb
\rho_{\rm sw} (\theta) = \rho_{\rm d}\frac{\gamma + 1}{\gamma - 1} \left(\frac{v_{\rm d}}{v_{\rm w}}\right)^2\frac{1}{(1 - \lambda)    r^{2}(\theta)}.
\label{rho2}
\eqe
As $\rho_{\rm sw} \propto \rho_{\rm d}$, changes in the density of the disk fluid will  directly affect the thermal bremsstrahlung power from the shocked stellar wind (see Sec.~\ref{sec:power}). The density of the shocked wind depends also on the thermal pressure of the disk, not only in a direct way through the $(1-\lambda)^{-1}$ term, but also indirectly through the shape of the shock surface $r(\theta)$ (see also Fig.~\ref{fig:r}). In particular, the density should increase for a higher $\lambda$ value for all angles $\theta$, while at the standoff-distance, i.e. for $\theta=0$, it scales as $\rho_{\rm sw}(0) \propto (1-\lambda)^{-1}$.

The normal components of the velocities across the shock front are related as $v_{\rm sw\bot} = \rho_{\rm w}v_{\rm w\bot}/\rho_{\rm sw}$,
where
\eqb
v_{\rm w \bot}(\theta)\equiv \bm{v_{\rm w}}\cdot \bm{\hat{n}}_{\bot} = v_{\rm w}\left(-\sin\theta \, A_{\rm z}(\theta) + \cos\theta \, A_\varpi(\theta)\right)
\label{v2norm}.
\eqe
The thermal pressure and temperature of the shocked region are written as
\eqb
\label{P2}
P_{\rm sw}(\theta) & = & \frac{2}{\gamma+1} \rho_{\rm w}v^2_{\rm w\bot}(\theta) \\
T_{\rm sw}(\theta) & = & \frac{(\gamma-1)}{(\gamma+1)^2}\frac{\mpr}{k_{\rm B}}v^2_{\rm w\bot}(\theta),
\label{T2}
\eqe
where $k_{\rm B}$ and $\mpr$ are the Boltzmann constant and the mass of the proton, respectively. From eq.~(\ref{T2}) we also retrieve the well-known result that the average thermal energy of the post-shock region,  $(3/2)k_{\rm B}T_{\rm sw}$,  is a fraction of the wind kinetic energy, $(3/32)\mpr v_{\rm w}^2$, where we used $\gamma=5/3$. The temperature of the shocked wind at $\theta=0$ is independent of the disk thermal pressure, i.e. $T_{\rm sw}(0)= (\gamma-1)\mpr v^2_{\rm w}/(\gamma+1)^2 k_{\rm B}$. This can be understood as follows: at the shock front, $P_{\rm sw} \propto (1-\lambda)^{-1}$ is determined by balancing the pressures while it was shown that $\rho_{\rm sw}(0)$ also scales as $(1-\lambda)^{-1}$. From the ideal gas equation of state, it directly follows that $T_{\rm sw}(0) \propto P_{\rm sw}(0)/\rho_{\rm sw}(0)\propto$ const.
\begin{figure*}
 \centering
\includegraphics[width=0.33\textwidth]{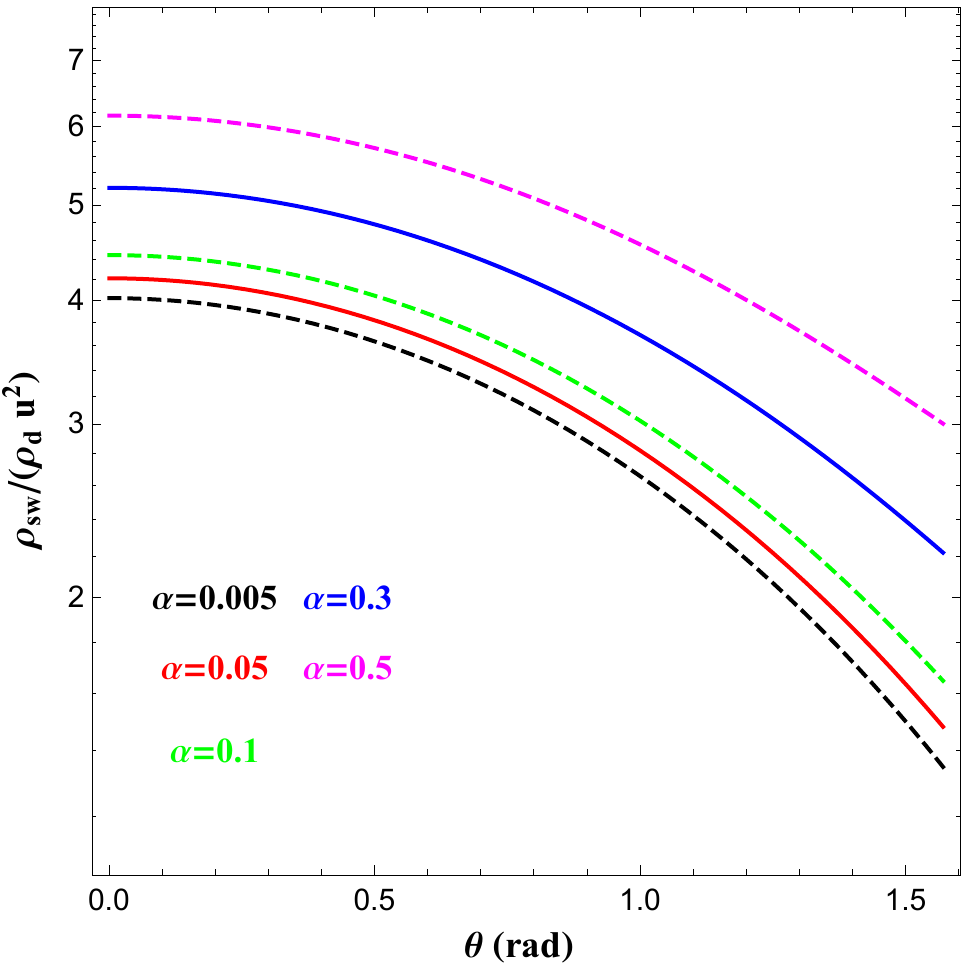}
\includegraphics[width=0.33\textwidth]{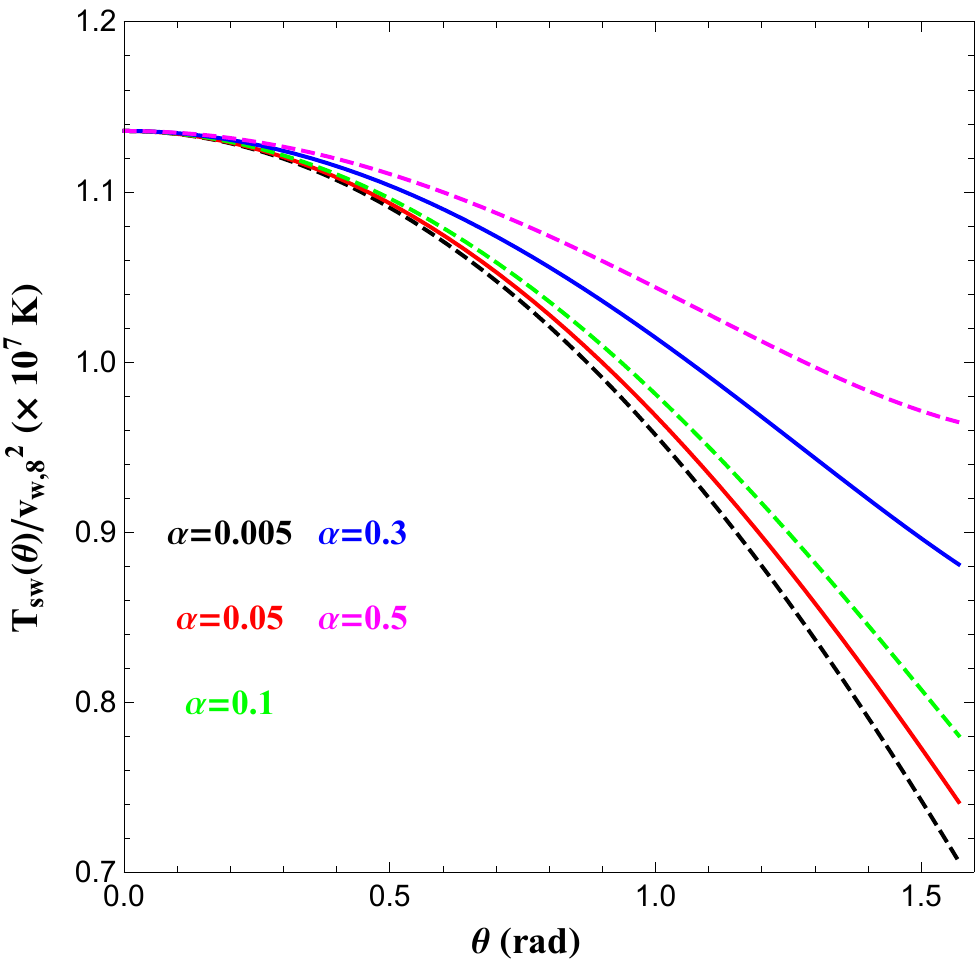}
\includegraphics[width=0.33\textwidth]{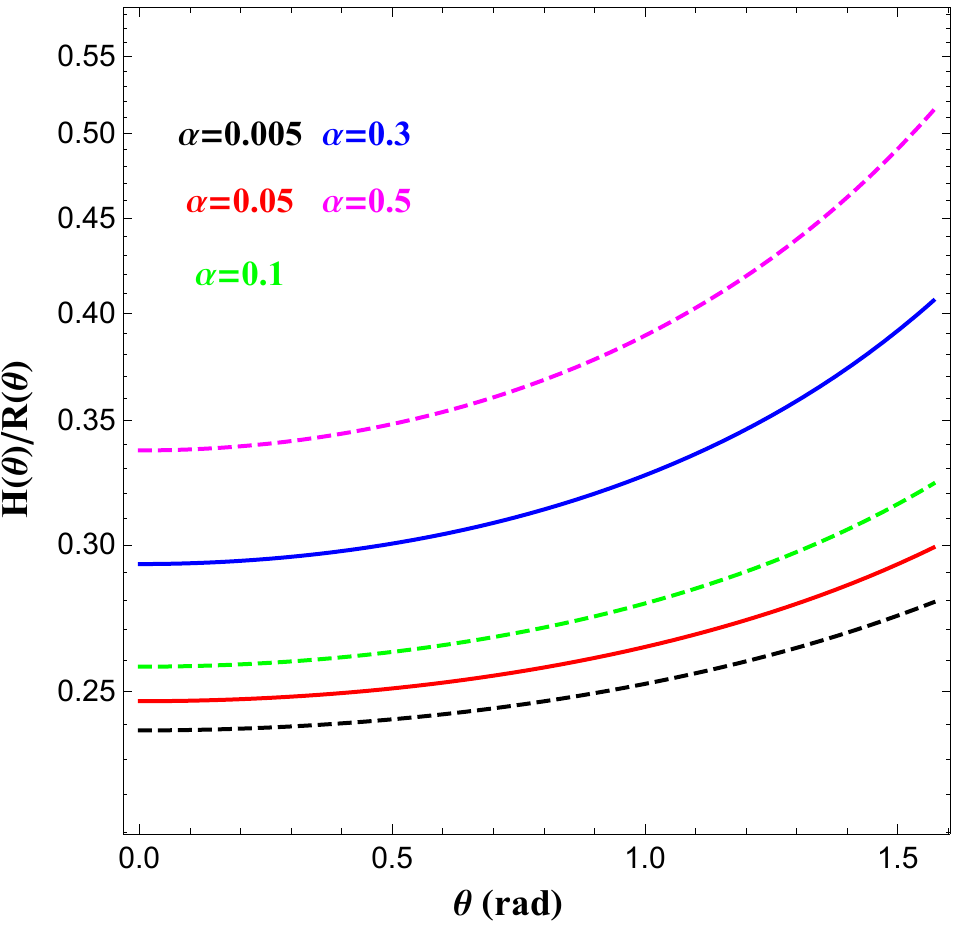}
\caption{Plot of the density ratio $\rho_{\rm sw}/\rho_{\rm d} u^2$ (left panel), the temperature $T_{\rm sw}$ (middle panel), and the thickness $H$ in units of $R$ of the shocked stellar wind shell (right panel) as a function of the angle $\theta$ for the $\alpha$ values marked on the plot. Here, $u=v_{\rm d}/v_{\rm w}$ and $v_{\rm w, 8} = v_{\rm w} / (10^{8}$ cm s$^{-1})$.}
\label{fig:dens_temp_thick}
\end{figure*}

The thickness of the shocked region $H$ can be estimated using the relation $H(\theta) \approx \sigma(\theta)/ \rho_{\rm sw}(\theta)$, where $\sigma$
is the  mass surface density of the shocked wind. This is defined as $\sigma = \Phi^2_{\rm m}/ R\sin\theta \, \Phi_{\rm t}$ and is given by
\eqb
\sigma(\theta) = \sigma_{0}\frac{g^2_\lambda(\theta)}{2 r(\theta) (1-\lambda) \sin\theta f_\lambda(\theta)}.
\label{sigma}
\eqe
where $\sigma_{0} \equiv R_0 \rho_{\rm d}$ and
\eqb
g_\lambda(\theta) = 2\frac{v_{\rm d}}{v_{\rm w}}(1-\cos\theta) + r^2\sin^2\theta (1-\lambda).
\label{g}
\eqe
By combining eqs.~(\ref{rho2}) and (\ref{sigma}) the thickness is found to be
\eqb
\frac{H(\theta)}{R(\theta)} = \frac{1}{2}\frac{\gamma-1}{\gamma+1}\left(\frac{v_{\rm w}}{v_{\rm d}}\right)^2\frac{g^2_\lambda(\theta)}{f_\lambda(\theta) \sin\theta}
\label{H}
\eqe

Figure~\ref{fig:dens_temp_thick} summarizes the semi-analytical results on the properties of the shocked wind: the angular dependence of the density ratio $\rho_{\rm sw}/\rho_{\rm d}$, the temperature $T_{\rm sw}$ and the thickness $H$ of the shocked stellar wind shell up to $\pi/2$.
A few things that are worth commenting follow:
\begin{enumerate}
 \item A higher thermal pressure in the disk causes the shock front to displace closer towards the star (see Figs.~\ref{fig:surfaces} and \ref{fig:r}).
 Both the increase of $\lambda$ and the decrease in $r(\theta)$ point towards a higher density ratio of the shocked stellar wind to the disk, $\frac{\rho_{\rm sw}}{\rho_{d}}$, for all angles $\theta$. This effect becomes prominent for $\lambda \approx \alpha > 0.1$.
 \item For a given $\lambda$, the ratio $\rho_{\rm sw}/\rho_{\rm d}$ becomes maximum at $\theta=0$, i.e. at the shock front. Although $\rho_{\rm sw}$ is a decreasing  function of $\theta$, at $\theta\simeq \pi/2$, it is only a factor of $\sim 1.5$ lower than its maximum value (see also eq.~(\ref{denpi}) in Appendix~\ref{app1}). Thus, the shocked stellar wind up to that angle is expected to have a non-negligible contribution to the radiated power.
 \item The temperature of the shocked wind, which is determined by the normal component of the stellar wind velocity, has a very weak dependence  on $\theta$. For $\theta=0$ the temperature is independent from the thermal pressure of the disk (see discussion after
 eq.~(\ref{T2})). It can also be seen that the temperature gradient in the $\theta$ direction is smoothed out as the thermal pressure of the disk increases. For typical values of the wind velocity, we find the temperature to be in the $\sim 10^{7}$ K range ($\sim$keV band).
 \item  The width of the shocked region, which is bounded by the contact discontinuity and the stellar wind shock front, increases
 for higher $\lambda$ values, since the shock front moves closer to the star (see also point (i)). For $\lambda \leq 0.1$, we find that $H/R \ll 1$
 for all angles up to $\pi/2$, which justifies the use of the thin-shell approximation. This begins to break down for the case of a very hot disk, where 
 $H/R \sim 0.35-0.55$.
\end{enumerate}
\section{Numerical simulations}
\label{numerical}
To test the validity of the assumptions used to derive the semi-analytical expressions presented in Sec.~\ref{model}, numerical hydrodynamic simulations were employed. Here, we report on the numerical setup used for describing the accretion disk - stellar wind interactions (i.e., initial and boundary conditions) and present the technical details of the numerical code. A cylindrical coordinate system was used in all simulations. Assuming axisymmetry, the simulations were performed in two dimensions and in the rest frame of the star, which   was assumed to be located at a distance $z_*$ along the symmetry $z$-axis.
\subsection{Numerical setup}
\label{sec:setup}
Simulations were carried out for $\lambda = 0.005, 0.1, 0.17, 0.23, 0.29$, and $0.35$ (equivalently for $\alpha=0.005, 0.1, 0.2, 0.3, 0.4$, and $0.5$).
The parameter values used as initial conditions in the numerical simulations are summarized in Table~\ref{tab0}.
All simulations were allowed to reach a steady state after an initial period of adjustment (for details, see Section \ref{sec:technical}).
In order to minimize the duration of the adjustment period, the simulations were optimized as described below:
  \begin{enumerate}
\item A radial stellar wind was continuously injected into the grid from a spherical surface of radius $R_{\rm inj}$ centered at $z_*$. For a wind velocity $v_{\rm w}$, the wind number density at the injection point was determined by $n_{\rm w} = \dot{M}_{\rm w}/(4\pi R_{\rm inj}^2 v_{\rm w} \mpr)$. The wind thermal pressure was chosen to  be a small fraction of the wind's ram pressure, namely
$P_{\rm w, th} = \alpha_{\rm w} P_{\rm w} = \alpha _{\rm w} n_{\rm w} \mpr v_{\rm w}^2$ with $\alpha_{\rm w}=10^{-3}$ or, $P_{\rm w, th} = 10^{-8} n_{\rm w} \mpr c^2$. Such low values of $P_{\rm w, th}$ were necessary for avoiding any unwanted wind acceleration.
\item A uniform disk medium was continuously injected from the $z=0$ boundary with velocity $v_{\rm d}$, number density $n_{\rm d}$ and thermal pressure $P_{\rm th} = \alpha n_{\rm d} \mpr  v_{\rm d}^2$.
\item Initially, the grid of the simulation box was filled with a uniform disk medium moving along the $z$ direction and having the same properties as these of the fluid that was being injected at the boundary $z=0$. The duration of the adjustment period is reduced with the particular setup. Alternatively, one could initially fill the grid of the simulation box with the stellar wind medium, assuming a $r^{-2}$ dependence of its properties with respect to the center of the star. It was verified that neither the results nor the duration of the adjustment period would change significantly in this case.
\item Reflection boundary conditions at the $z-$axis (inner radial boundary) and outflow boundary conditions at both (radial and axial) outer boundaries were imposed. The outer boundaries had to be placed sufficiently far away from the star to avoid the propagation of numerical artifacts produced at those boundaries towards the wind-disk interaction region. Typically, the outer radial and axial boundaries were, respectively, located at $1.2\times 10^{15}$ cm and $2.4\times10^{15}$ cm.
\end{enumerate}

\begin{table}
 \centering
 \caption{Values of the fixed parameters used in the numerical simulations.}
 \begin{threeparttable}
 \begin{tabular}{ccc}
  \hline
Parameter & Symbol & Reference value \\
\hline
 Stellar wind velocity (cm s$^{-1}$) & $v_{\rm w}$ & $10^8$\\
Mass-loss rate ($M_\odot\, {\rm yr}^{-1}$) & $\dot{M}_{\rm w}$ & $10^{-7}$\\
 Wind number density (cm$^{-3}$) &$n_{\rm w} $& $480\times10^{4}$\\
 Disk velocity\tnote{a} \,\,  (cm s$^{-1}$) & $v_{\rm d}$ & $8\times 10^8$\\
 Wind injection radius (cm) & $R_{\rm inj}$ & $2.5\times10^{13}$\\
 Ratio of thermal to ram disk pressure\tnote{b} & $\alpha$ & $0.005$, $0.1$, $0.2$ $0.3$, $0.4$, $0.5$\\
\hline
 \end{tabular}
 \tnote{a} Measured in the rest frame of the star.\\
 \tnote{b} The respective Mach number of the disk fluid with respect to the shock for the listed $\alpha$ values are $11$, $2.4$, $1.7$, $1.4$, $1.2$, and $1.1$ respectively.
 \end{threeparttable}
\label{tab0}
\end{table}

\subsection{Technical details}
\label{sec:technical}
\begin{figure}
 \centering
 \includegraphics[width=0.42\textwidth]{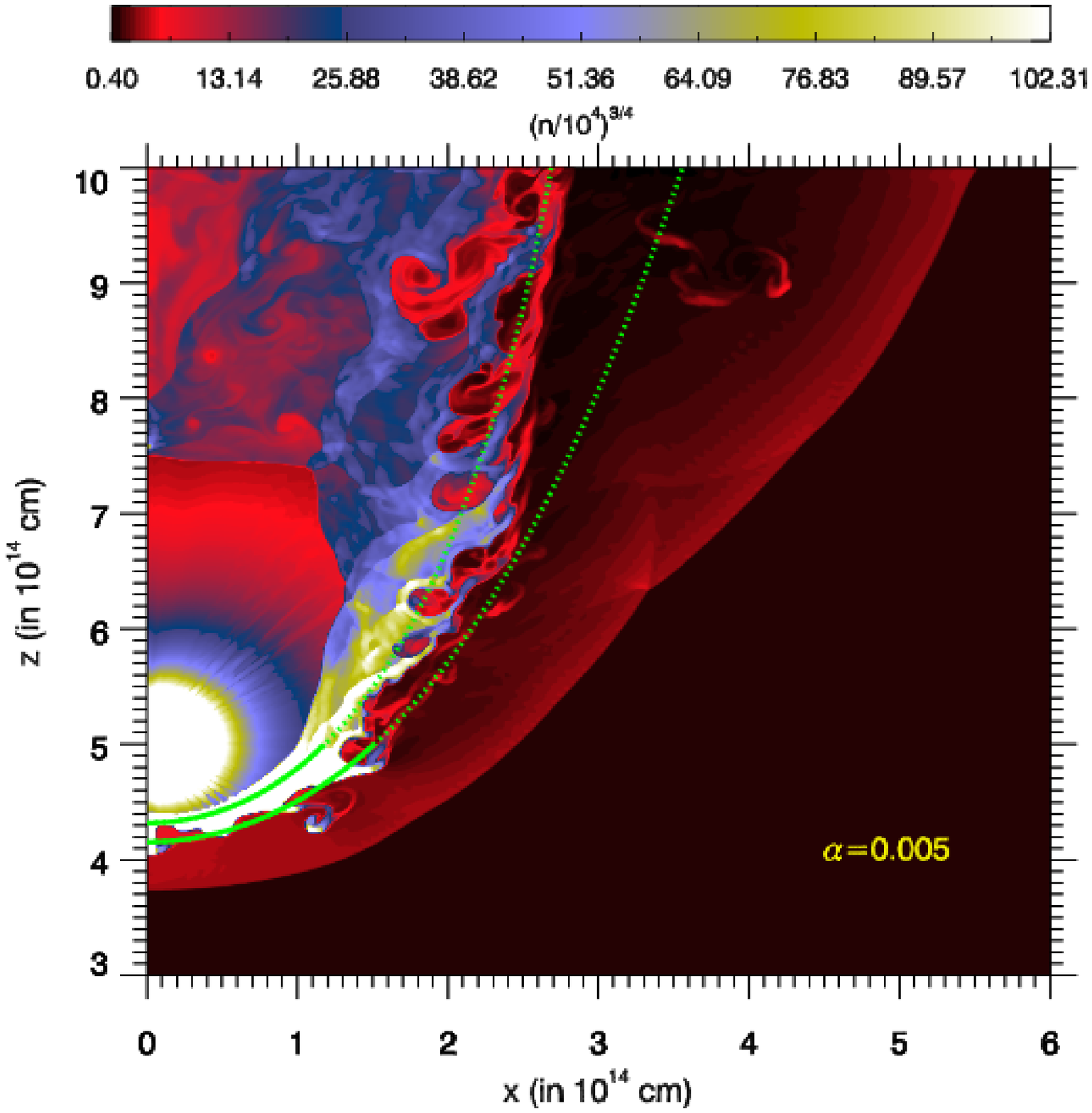}
 \includegraphics[width=0.42\textwidth]{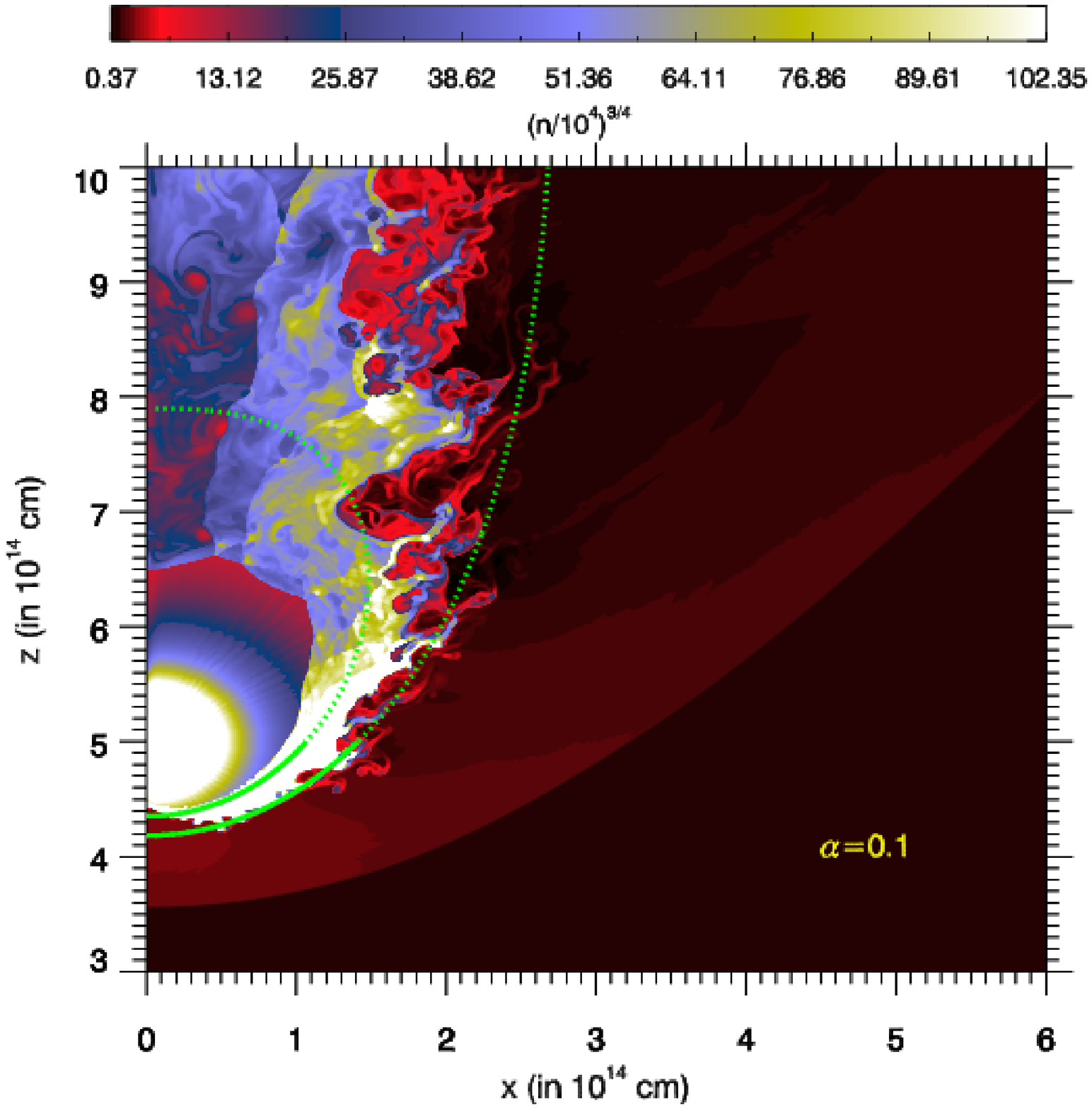}
 \includegraphics[width=0.42\textwidth]{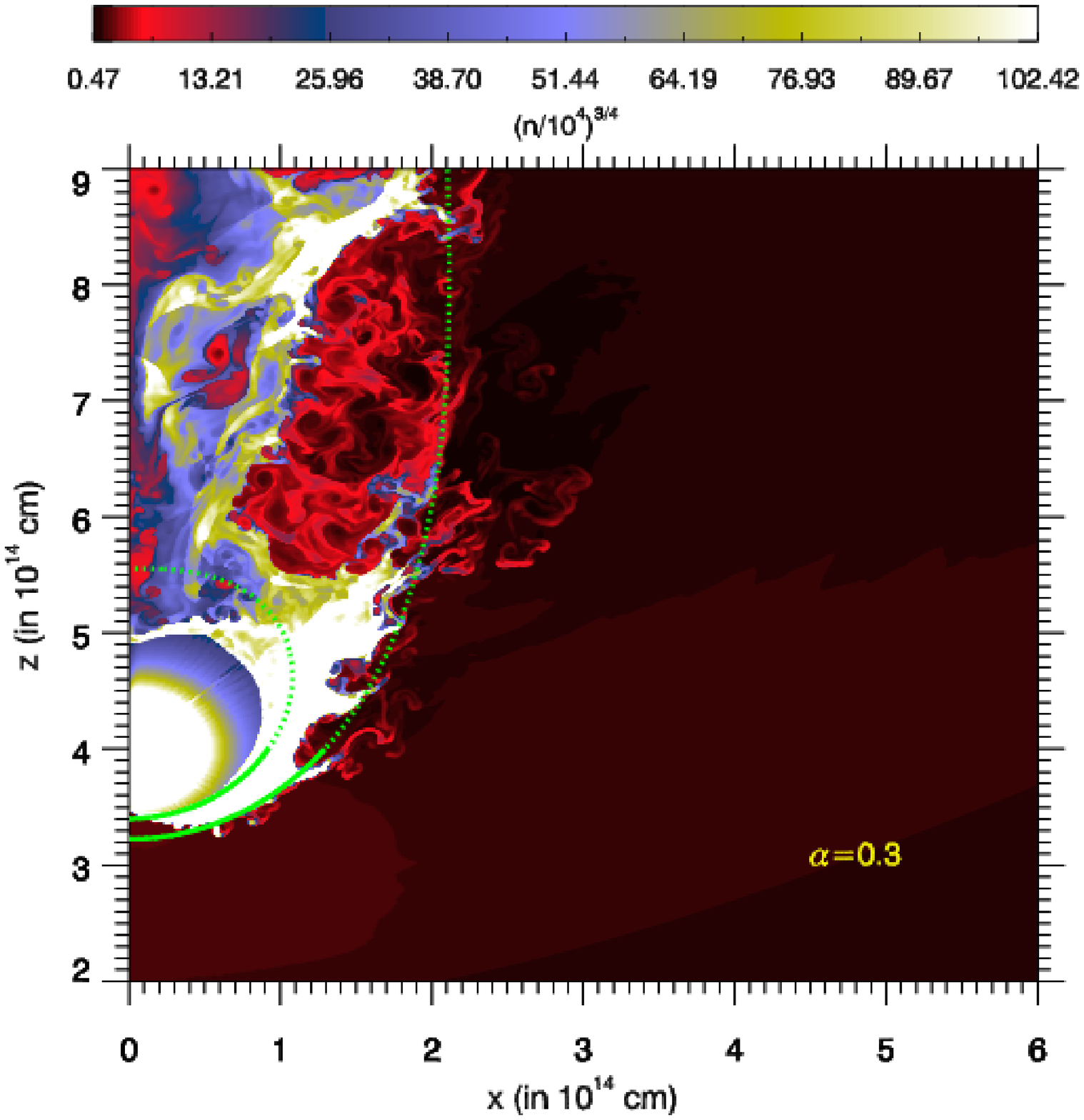}
 \caption{Density maps as obtained from the hydrodynamic simulations for $\alpha=0.005$, $\alpha=0.1$, and $\alpha=0.3$. The semi-analytical results for the position of the termination shock $R(\theta)$ and the contact discontinuity $R(\theta)+H(\theta)$ are shown as thick green lines for $\theta \le \pi/2$ (solid lines) and $\theta > \pi/2$ (dotted lines). The actual value of the density (in units of $10^{4}{\rm cm}^{-3}$) can be read from the colorbar after raising the listed values to the power of 4/3.}
\label{curvecomparison}
\end{figure}

 High-resolution simulations  of the wind-disk interaction were performed using the relativistic hydrodynamics code \textsc{mrgenesis} \citep{mimica2009}.
 A uniform numerical grid with a typical spacing $\sim 6.67\times 10^{11}$ cm was used. In terms of numerical resolution, this corresponds to a resolution of $1800$ points in the radial and $3600$ points in the axial direction. \textsc{mrgenesis} uses a third order total variation diminishing Runge-Kutta scheme {\citep{ShuOsher1988}} for the time integration and the piecewise-parabolic method {\citep[PPM;][]{ColellaWoodward1984}} for the spatial interpolation. The intercell fluxes are computed with the Marquina flux formula {\citep{Donat1998}}.  The fluids are assumed to obey the ideal gas equation of state with the adiabatic index $5/3$. The code is parallelized using MPI\footnote{Message Passing Interface, \url{http://www.mpi-forum.org}. }.   All simulations were performed on the Tirant supercomputer\footnote{More information (in Spanish): \url{http://www.uv.es/siuv/cas/zcalculo/res/descripcion.wiki} at the University of Valencia.
}.
 In all cases, the simulation time was set to $\sim 10^4$ hours, while the typical adjustment period lasted $\sim 10^3$ hours. Snapshots of the grid state were saved every $\simeq 20$~hours, allowing us to follow the temporal evolution of the wind-disk interaction. Each snapshot contained the density, pressure and velocity maps (for density maps, see Fig.~\ref{curvecomparison}) that were used  to compute the free-free emissivity (e.g. Fig.~\ref{fig:jff}).

\subsection{Comparison}
We first performed a qualitative comparison between the results of the semi-analytical model and the numerical simulations regarding the shape of the shock surface and proceeded with a quantitative comparison of the properties of the shocked stellar wind, such as temperature and density angular profiles. For this purpose, after verifying that a steady state has been reached, the simulation snapshots for different $\lambda$ values were post-processed in order to obtain the following: (i) the characteristic distances $R_0$ and $R_1$, (ii) the temperature and density profiles of the shocked stellar wind shell, and
(iii) the thermal bremsstrahlung power emitted from the shocked stellar wind.

\subsubsection{Geometrical characteristics}
\label{sec:geometry}
Figure~\ref{curvecomparison} presents two-dimensional density maps as obtained from the simulations for $\alpha=0.005, 0.1$ and $0.3$. In the first two cases the star is positioned at $(x_{\star},z_{\star})=(0,5)$, while in the third case, its location is $(0,4)$. The density values (in units of $10^4$ cm$^{-3}$) can be read from the colour tables included in each panel. The semi-analytical results for $R$ and $R+H$ as a function of the angle $\theta$ are overplotted with thick yellow lines. The results for $\theta \le \pi/2$ are shown with solid lines, whereas dotted lines are used for larger angles.
Inspection of Fig.~\ref{curvecomparison} shows that the derived curves straddle along the interfaces fairly well up to an angle of $\pi/2$. At larger angles, the results start to diverge from the numerical ones; hence, the reasoning for the $\theta$ plotting range in the figures of Sec.~\ref{model}.  Furthermore, the numerical results verify the assumption of a homogeneous post-shock region, at least up to $\pi/2$. The main reason for the deviation between the semi-analytical and numerical results is the growth of Kelvin-Helmholtz instabilities that cannot be treated within the adopted analytical framework. These take place at the interface separating the shocked disk fluid and the shocked stellar wind (i.e. the contact discontinuity).  The mixing of the fluids spreads over a larger volume as the thermal pressure of the disk increases, while it leads to more turbulent flows. 

Regardless, the semi-analytical model captures the rough features of the shocked wind shell. The only appreciable difference between the numerics and analytics for $\theta \lesssim \pi/2$ is identified close to the symmetry axis (see bottom panels in Fig.~\ref{curvecomparison}) and is a numerical artifact caused by the reflection boundary along the symmetry axis. As will describe below, this deviation is important when comparing geometrical characteristics, but for the comparison of volume-integrated quantities, such as the thermal radiated power, it does not have a significant effect. Nevertheless, this deviation tends to diminish as the resolution increases.

The standoff-distance, $R_0$, marks the starting point of the analytical model and is the best estimate for the location of the termination shock at $\theta=0$. Similarly, a characteristic distance $R_1$ can be defined at $\theta\approx \pi$ (see Sec.~\ref{model}). Both are found by matching pressures on either side of their respective interface. These values were determined by first creating a pressure map, similar to the density maps in Fig.~\ref{curvecomparison}, of a snapshot taken after the simulation has reached a steady-state. The first non-negligible jumps in the pressure found at $z<z_{\star}$ and $z>z_{\star}$ represent the locations of $R_0$ and $R_1$, respectively. The results are summarized in Table~\ref{r0table}. Overall, a good agreement is found: the distances determined by the simulations scale with $\alpha$ as predicted by the analytics (e.g. see eq.~(\ref{Ro})), while the fractional errors are $\sim 20\%$ for all cases with $\alpha >0.1$. 
For colder disks ($\alpha \ll 0.1$), the original assumption, that the pressure of the post-shock wind is of the same order as the disk pressure, breaks down, thus preventing us from obtaining a reasonable value for $R_1$ (see also Table~\ref{r0table}). In the specific example of $\alpha = 0.005$, the numerical simulations show that $P_{\rm sw} \gg P_{\rm th}$.
\begin{figure*}
 \centering
\includegraphics[width=0.40\textwidth]{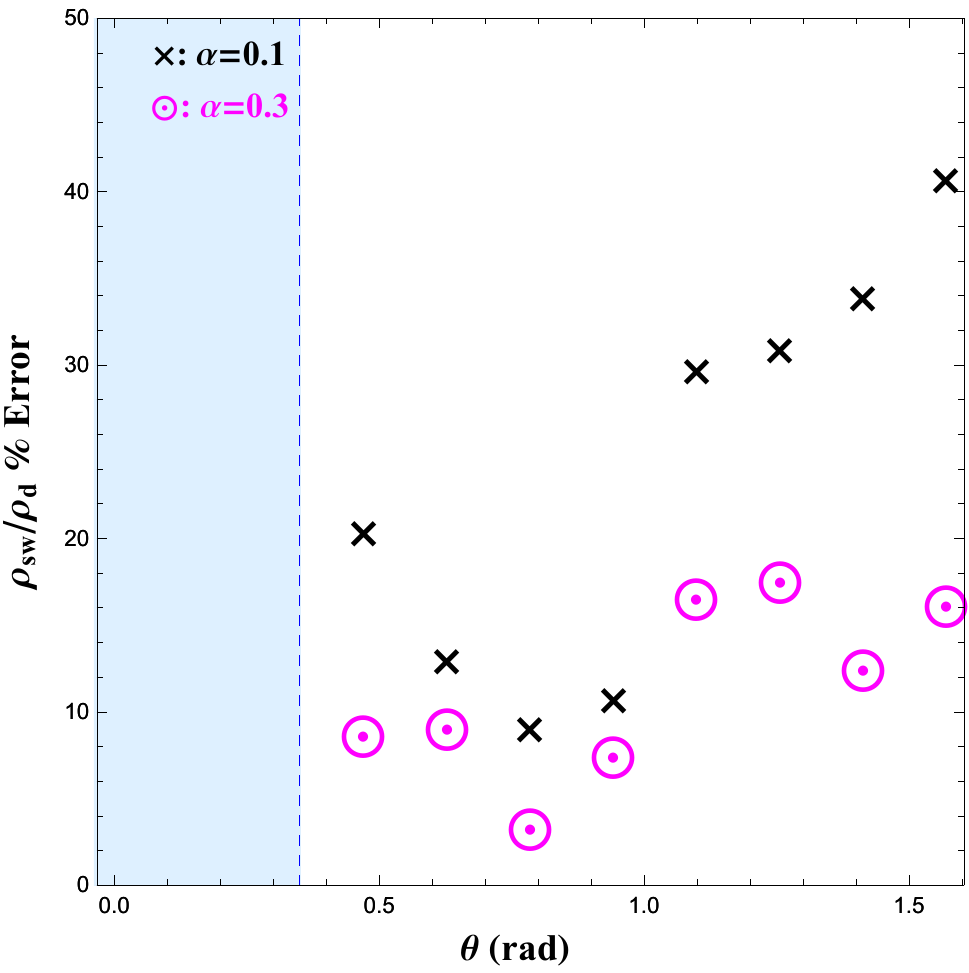}
\hspace{1in}
\includegraphics[width=0.40\textwidth]{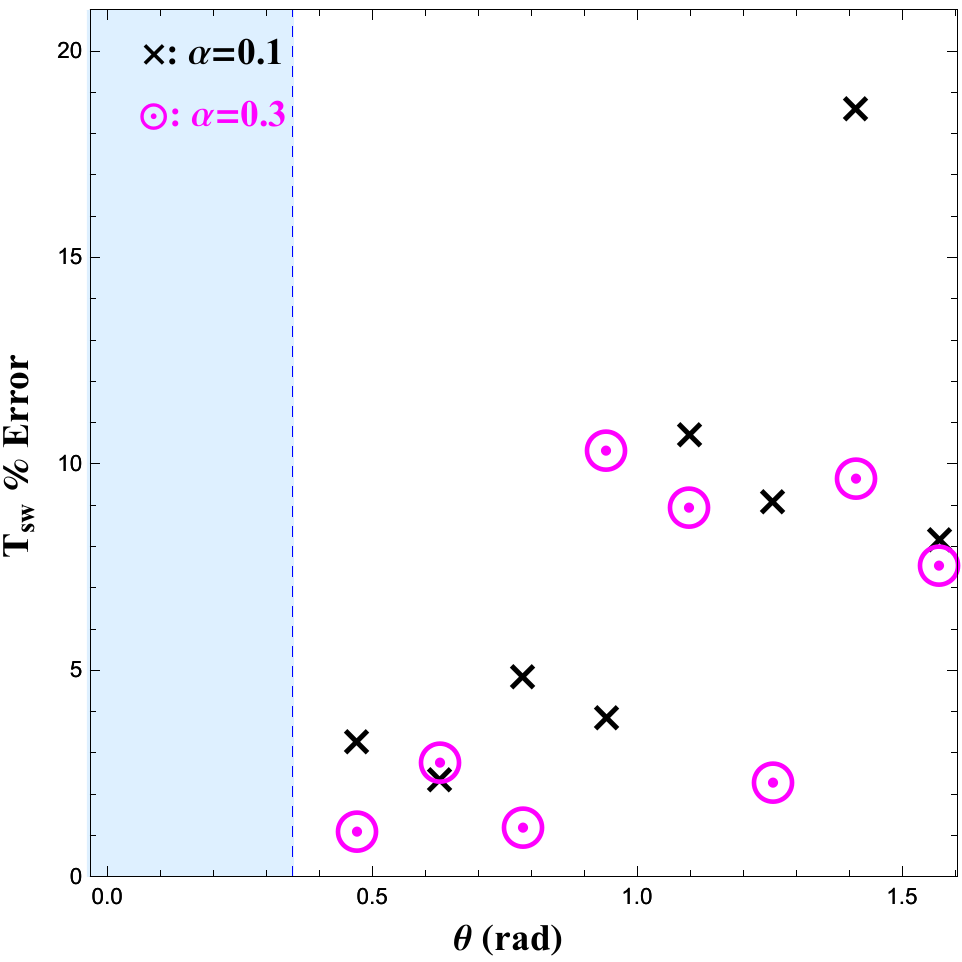}
\caption{Percentage error of the density ratio (left panel) and temperature of the shocked stellar wind (right panel) obtained from the simulations as compared with the theoretical (i.e. semi-analytical) expressions for different $\theta$ values. The shaded area denotes the angular region where numerical artifacts related to the inner axial boundary are non-negligible, thus not allowing for a proper comparison with the semi-analytical results.}
\label{fig:tempdencompare}
\end{figure*}

The quantitative differences found between the analytical and numerical values of $R_0$ are most likely caused by numerical artifacts at the inner axial boundary of the simulation. By further examining the $n_{\rm w}$ and $v_{\rm w}$ of the unshocked wind close to $\theta = 0$, within the region bounded by the shock front and the injection radius, it was found that $n_{\rm w}$ is twenty percent lower than the specified initial condition for $\alpha = 0.005$ while it is only ten percent lower for the remaining $\alpha$ values. 

The deviations in $R_1$, displayed in Table~\ref{r0table}, are much larger than those in $R_0$. Moreover, the distance $R_1$ oscillates during the simulation about a mean value but never reaches a stationary value, in contrast to $R_0$. Both results are related to the growth of Kelvin-Helmholtz instabilities within a significant fluid volume behind the star (see Fig.~\ref{curvecomparison}). On the one hand, the instabilities cause waves to propagate towards the axis (behind the star) upon which they are reflected. The reflected waves, in turn, make the shape of the shock surface far beyond the star to oscillate. On the other hand, the shocked disk medium behind the star is very complex due to Kelvin-Helmholtz instabilities, thus leading to large pressure changes; the pressure at e.g. $\theta=\pi/2$ can vary by as much as one order of magnitude which corresponds to small variations in $R_1$.

\begin{table}
\centering
\caption{Distance of the termination shock at $\theta=0$ and $\theta=\pi$ as derived from the simulation and the semi-analytical model. For $\alpha=0.005$, the model assumption required for the determination of $R_{1}$ breaks down (for details, see text).}
\begin{tabular}{cccccc}
\hline
$\alpha$ & $\lambda$ & \multicolumn{2}{c}{$R_0$ (in $10^{14}$cm)} &  \multicolumn{2}{c}{$R_1$ (in $10^{14}$cm)}\\
\hline
\multicolumn{2}{c}{\phantom{}} &  model &  simulations & model & simulations\\
\hline
  0.005 & $0.005$ & $0.68$ & $0.60$& $9.68$ & $2.52$  \\
  0.1   & $0.10$ & $0.65$ & $0.54$& 2.06 & 1.56\\
  0.2   & $0.17$ & $0.63$ & $0.52$ & $1.54$ &$1.20$\\ 
  0.3   & $0.23$ & $0.60$ & $0.51$&  1.25 &  0.95\\
0.4 & 0.29 & 0.58 & 0.50 & 1.08 & 0.82 \\
  0.5   &  $0.35$ & $0.55$ & $0.48$& 0.93 & 0.77\\
  \hline
\end{tabular}
\label{r0table}
\end{table}

\subsubsection{Temperature and density profiles}
\label{sec:temp}
The density and temperature of the shocked stellar wind region as a function of $\theta$ are described by the analytical expressions (\ref{T2}) and (\ref{rho2}) presented in Sec.~\ref{model}. In order to compare the derived results with those from the simulations, density and temperature values were extracted from their respective two-dimensional maps for different $\theta$ values while moving along a curve that falls within the shell.
Based on the assumption that these quantities are radially independent (refer to the simulation snapshots in Fig.~\ref{curvecomparison} for a visual confirmation), any choice of curve that overlaps with the shocked shell is acceptable. Here, a curve was chosen with a radial distance of $r(\theta) + \frac{H(\theta)}{4 R_0}$ from the star. 

The main objective in the comparison of the temperature and density profiles is to determine the deviation between the semi-analytical and numerical results. For this procedure, the percentage error was computed while adopting the semi-analytical results as the exact values. The results of the comparison for $\alpha = 0.1$ and $0.3$ are displayed in Fig.~\ref{fig:tempdencompare}. In each plot, angles close to zero, represented by the shaded region in the figure, were excluded in the comparison for reasons explained in the previous section. The results for the temperature (right panel in Fig.~\ref{fig:tempdencompare}) are in good agreement with the semi-analytical model and fall within $15\%$ of the predicted values. As for the angular profile of the density ratio (left panel in Fig.~\ref{fig:tempdencompare}), a good agreement is found for both $\alpha =0.3$ and $0.1$ with most errors falling within $35\%$ of the theoretical values. However, there is a slight deviation found in the $\alpha = 0.1$ case for 
large angles. This deviation is reasonable for angles close to $\pi/2$, for this is the point at which the model tends to break down. 

\subsubsection{Thermal bremsstrahlung luminosity}
\label{sec:power}
As the stellar wind is shocked by the interaction with the accretion disk, the shocked gas begins to cool via thermal bremsstrahlung emission.
The total radiated thermal luminosity from the shocked shell is given by
\eqb
L_{\rm shell} = 2\pi \int_{0}^{\pi/2} \sin\theta \, {\rm d}\theta \int_{R(\theta)}^{R_{\max}(\theta)}\!\!\!{\rm d}R' R'^2 \frac{{\rm d}^2W\left(\theta\right)}{{\rm d}t {\rm d}V},
\label{lum}
\eqe
where $R_{\max}(\theta)\equiv R(\theta)+H(\theta)$, ${\rm d}^2W/{\rm d}t {\rm d}V$ is the free-free emissivity, and
the integration is performed up to $\theta= \pi/2$ where the agreement between the semi-analytical model and the simulations is  good. The above expression also implies that there is no gradient of the temperature and density in the radial direction within the shell, namely that the shocked fluid in the shell is homogeneous. This is verified by the simulation results (see Fig.~\ref{curvecomparison}). For the case of a wind composed of pure hydrogen, analytical expressions for the free-free emissivity are available
(see e.g. eq.~(5.15b) in \cite{rybicki_lightman1986}). In the more realistic scenario of a stellar wind with solar metallicity, appropriate tables for the free-free cooling function were used \citep{sutherland1993}.

Before performing any computation of the radiated power produced by the shocked wind, it is important to remove any contributions from the unshocked wind and disk. The emission from the unshocked disk, in particular,  is expected to be non-negligible for large $\alpha$ values and is considered to be a background emission for the purposes of the present study. To remove these `unwanted' regions from the simulation snapshots, certain cuts in the temperature maps were made by setting the temperature of specific data points to zero. Throughout every simulation, it was found that the shocked wind never reaches temperatures below $10^{6}$ K. Cutting any data point with a temperature below this value allowed for the removal of unwanted contributions from the cool, unshocked wind. A cut is also made for regions with temperatures above $10^{8.5}$ K, as this is the maximum temperature value included in the cooling factor tables \citep{sutherland1993}.

The total bremsstrahlung power produced by various regions within the simulation is computed for the cases of pure hydrogen and solar metallicities by converting the general form of the double integral in~eq.~(\ref{lum}) into a double, discrete sum, namely
\eqb
L = 2 \pi \sum_{z} \sum_{x} x \frac{{\rm d}W(x, z)}{{\rm d}t {\rm d}V} \Delta z \Delta x
\label{lumsim}
\eqe
where $\Delta x$ and $\Delta z$ are the spacing between each data point in the $x$ and $z$ directions, respectively. The total radiated power produced by the shell is determined by eq.~(\ref{lumsim}) while limiting the angle of integration up to $\theta = \pi/2$.
\begin{figure}
 \centering
\includegraphics[width=0.48\textwidth]{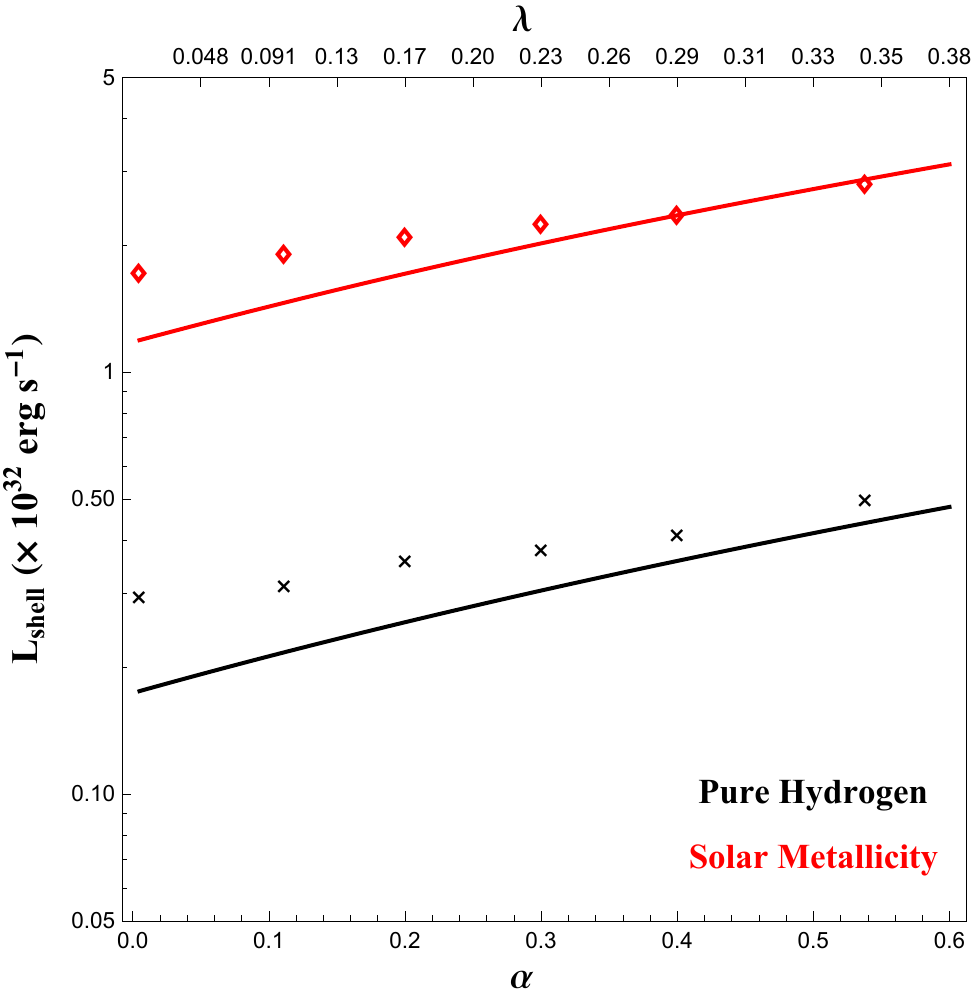}
\caption{Logarithmic plot of the thermal bremsstrahlung luminosity emitted from the shocked stellar wind shell as function of $\alpha$. The results obtained assuming gas with solar metallicity and pure hydrogen are plotted with red and black lines, respectively. The results from simulations for various $\alpha$ values are overplotted with symbols.}
\label{fig:lum}
\end{figure}
The results are presented in Fig.~\ref{fig:lum} where the simulation markers are representative of the six $\alpha$ values given at the beginning of Sec.~\ref{numerical}. The results obtained assuming gas with solar metallicity and pure hydrogen are plotted with red and black lines, respectively. Inspection of the figure shows a good agreement between the results of the semi-analytical model and the simulations for both gas compositions considered. The outcome of the comparison was not trivial, given that the radiated power is a convolution of the gas temperature, density and emitting volume. 
When solar metallicity is taken into account, an increase of the radiated power is found, for all $\alpha$ values, compared to the case of pure hydrogen. The effect becomes more prominent for hotter disks and is of importance for the application to Sgr~A*, as will be shown next in Sec.~\ref{sec:application}.

For fluids composed of pure hydrogen, in particular, the good agreement between the results allows us to use the semi-analytical model in order to derive scalings of the bremsstrahlung luminosity on parameters, such as the wind velocity and disk thermal pressure.  The bolometric bremsstrahlung power of the shell may be written as
\eqb
L_{\rm shell} \propto \dot{M}_{\rm w}^{3/2} n_{\rm d}^{1/2} v_{\rm w}^{-3/2} v_{\rm d} (0.3 + 0.47 \, \alpha^{1.41}).
\label{scale}
\eqe
The appropriate scaling is determined by first noting that the free-free emissivity, given in eq.~(\ref{lum}), scales as $T_{\rm sw}^{1/2}$ and $n_{\rm sw}^{2}$ (see also eq.~(5.14a) in \cite{rybicki_lightman1986}), while the total luminosity scales as $R_{0}^{3}$. By use of eqs.~(\ref{Ro}), (\ref{rho2}), and (\ref{T2}) one may accurately obtain the appropriate scalings for $\dot{M}_{\rm w}$, $n_{\rm d}$, $v_{\rm w}$, and $v_{\rm d}$. To obtain the appropriate scaling for $\alpha$ (i.e. last term in the right-hand side of eq.~(\ref{scale})), a non-linear fit was performed on the black points in Fig.~\ref{fig:lum}.

Similar methods can be used to determine the appropriate parameter scalings for a stellar wind composition of solar metallicities. For this case, the luminosity has the same scaling for $\dot{M}_{\rm w}$ and $n_{\rm d}$ as in eq.~(\ref{scale}) while the dependence on $\alpha$ is $(1.83 + 2.2\, \alpha^{1.18})$. The $\alpha$ scaling is made by performing a non-linear fit to the red points in Fig.~\ref{fig:lum}.
\subsection{The role of Kelvin-Helmholtz instabilities on the bremsstrahlung luminosity}
\label{sec:khinstabilities}
\begin{figure}
 \centering
\includegraphics[height=8cm, width=8cm]{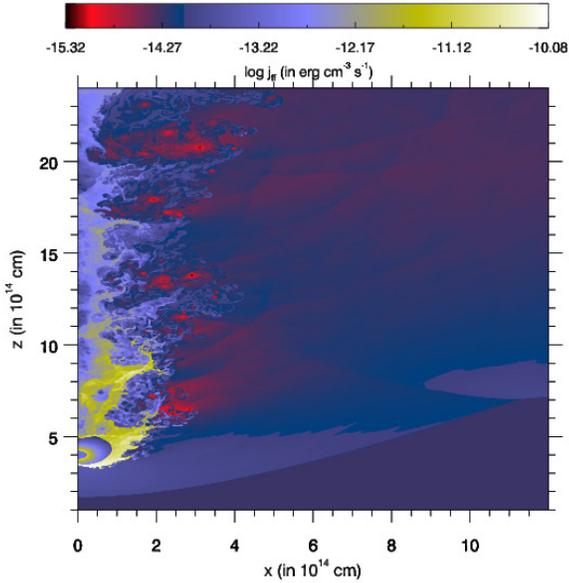}
\caption{Free-free emissivity map as obtained from the simulation for $\alpha=0.3$, assuming gas composed of pure hydrogen. The star is located at $(x_\star,z_\star)=(0,4)$.}
\label{fig:jff}
\end{figure}

A result that could not be predicted from the semi-analytical model presented in Sec.~\ref{model} is the efficient mixing of the disk and wind fluids
due to Kelvin-Helmholtz instabilities. Although the density of the gas is not as high as in the shocked shell region, the occupied volume by the mixed fluids
is considerably larger. This is illustrated in Fig.~\ref{fig:jff} where a large scale, two-dimensional map of the free-free emissivity, in the case of pure hydrogen, is plotted for a hot disk with $\alpha=0.3$. Although the details of the turbulent structures are not visible, Fig.~\ref{fig:jff} clearly shows that the emissivity from the mixed fluids at distances of $14\times 10^{14}$ cm away from the star is non-negligible.
\begin{table}
\centering
\caption{Total luminosity produced from simulations including contributions from the long tail of mixed fluids behind the star.}
\begin{tabular}{cccc}
\hline
	&           & Pure Hydrogen & Solar Metallicity  \\
\hline
$\alpha$& $\lambda$ &$L$ ($\times 10^{32}$ erg s$^{-1}$) &  $L$ ($\times 10^{33}$ erg s$^{-1}$) \\
\hline
0.005 & $0.005$ & $0.736$ & $0.511$ \\
0.1 & $0.1$ & $2.008$ & $1.168$ \\
0.2 & $0.17$ & $3.017$ & $1.541$\\
0.3 & $0.23$ & $3.752$ & $1.759$\\
0.4 & $0.29$ & $5.292$ & $2.424$\\
0.5 & $0.35$ & $6.203$ & $2.815$\\
\hline
\end{tabular}
\label{totlum}
\end{table}
To better assess the role of Kelvin-Helmholtz instabilities on the bremsstrahlung luminosity, the radiated power was computed by use of eq.~(\ref{lumsim}) for a larger volume within the simulation. This volume included the shocked shell region for $\theta > \pi/2$ and the regions extending beyond $R_1$, which stretches anywhere between $7\times 10^{14} - 2\times 10^{15}$~cm above $z_*$.
The results are summarized in Table~\ref{totlum} and demonstrate that the mixing region produces a significant contribution to the luminosity of the entire simulation. This is of the same order or even larger than that produced by the shocked shell as compared with Fig.~\ref{fig:lum}.

\begin{figure}
 \centering
\includegraphics[width=0.48\textwidth]{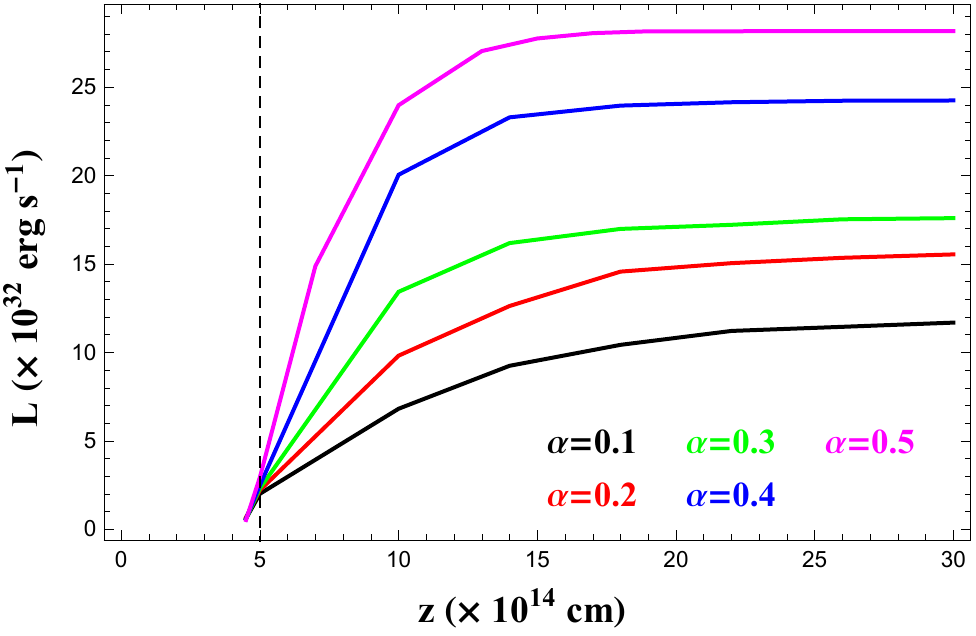}
\caption{Cumulative bremsstrahlung luminosity, $L(\le z)$, calculated assuming solar metallicity, for the $\alpha$ values marked on the plot. The vertical line indicate the position of the star, $z_* = 5\times10^{14}$ cm, in the simulation runs.}
\label{fig:cumL}
\end{figure}

The results listed in Table~\ref{totlum} also raise the question of whether the size of the simulation box is sufficiently large enough to capture all the radiated power produced by the mixed fluids. This issue can be addressed by calculating the cumulative luminosity $L(\le z)\equiv \int^{z} {\rm d}z^\prime  L(z^\prime)$. This is presented in  Fig.~\ref{fig:cumL} for the five $\alpha$ values listed at the beginning of Sec.~\ref{sec:setup} while assuming solar metallicity. The vertical line indicates the position of the star, $z_*= 5\times10^{14}$ cm, in the simulation runs. In both cases, it was found that the cumulative luminosity begins to converge to a final value at $z \approx 2.5 z_* \approx 12\times10^{14}$ cm. However, the gradient along the symmetry axis, i.e. ${\rm d} L/{\rm d}z$, is much steeper in hotter disk cases. The increase in $\alpha$ also causes an increase in the total luminosity of the simulation snapshots; this is found to scale as $\alpha^{0.56}$.

\section{Application to Sgr~A*}
\label{sec:application}
Sgr A*, the compact radio source associated with the supermassive black hole in our Galactic center, spends most of its time in a low-luminosity, X-ray emission state with an absorption corrected luminosity of $\approx 2.4 \times10^{33}$ erg s$^{-1}$ \citep{baganoff2003, genzel2010}. However, X-ray flares with durations ranging from minutes to hours and luminosities well above the quiescent emission of Sgr~A* are frequently observed \citep[e.g.][]{baganoff2001, nustar2014}. The physical cause of these rapid flares is not yet understood. Some of the hypothetical scenarios used to model these occurrences are an episodic outflow triggered by magnetic reconnection \citep{yuan2009}, enhanced mass accretion \citep{tagger2006}, or a possible hot spot in the accretion flow \citep{eden2010}.

Month-long X-ray flares are expected from the pericenter passage of certain S-stars as they precess through the accretion disk of Sgr A* \citep{giannios2013}. The present work will be oriented towards S2, whose upcoming pericenter passage will occur in 2018. This early, B-dwarf star is characterized by a mass-loss rate of $\dot{M}_{\rm w}\leq 3\times 10^{-7} M_\odot {\rm yr}^{-1}$, stellar wind velocities of $v_{\rm w} = 10^{8}$ cm s$^{-1}$, and a pericenter orbital velocity of $v_{\rm d} = 8\times 10^{8}$ cm s$^{-1}$. During the transit of the star through the disk, the star will encounter sections of the disk with different densities. In the following section, this change in density is mimicked by employing the use of a time-dependent density simulation, without adopting a specific model for the accretion disk.

\subsection{Mimicking the pericenter passage of S2}
As shown in Sec.~\ref{numerical}, a significant fraction of the bolometric bremsstrahlung emission originates from a long
tail of mixed fluids. For the astrophysical application we are interested in, it is not known {\sl a priori}
if the S2 pericenter passage lasts long enough as to allow for the full mixing of the fluids to distances far behind the star.
If $t_{\rm per}$ and $t_{\rm mix}$ denote the two characteristic timescales of the pericenter passage and mixing, the maximum bremsstrahlung luminosity is expected for the case $t_{\rm per} \gg t_{\rm mix}$. To address this issue, time-dependent simulations are required.

In what follows, the pericenter passage of the star is mimicked by varying the disk density used as an initial condition in the simulation.
The temporal dependence of the disk number density is modeled as
\eqb
\tilde{n}_{\rm d}(T) = 1 + \frac{9}{\cosh\left( \frac{T - 25 T_0}{3 T_0} \right)}
\eqe
where $T_0 = 7.5\times 10^{5}$ s and $\tilde{n}_{\rm d}=n_{\rm d}/(10^{4} {\rm cm}^{-3})$.
The density smoothly increases from its reference value at $T=0$, reaches a maximum value of $\tilde{n}_{\rm d}=10$ at $\sim 5200$~hr, and decreases to its initial value onwards. The density peaks on a timescale of $\sim 3 T_0 \sim 1$ month, which is comparable to the time it takes S2 to complete its pericenter passage. Due to the variable disk density, the surface of the termination shock oscillates by factor of 2 in distance, leading also to a variable bremsstrahlung emission. Interestingly, the region of mixed fluids, described in Sec.~\ref{sec:khinstabilities}, was quickly formed before the peak density was reached, while it was found to extend to a fairly large distance beyond the star. The inclusion of this tail region to the total luminosity of the system is, therefore, a valid assumption.

\begin{figure}
 \centering
\includegraphics[width=0.48\textwidth]{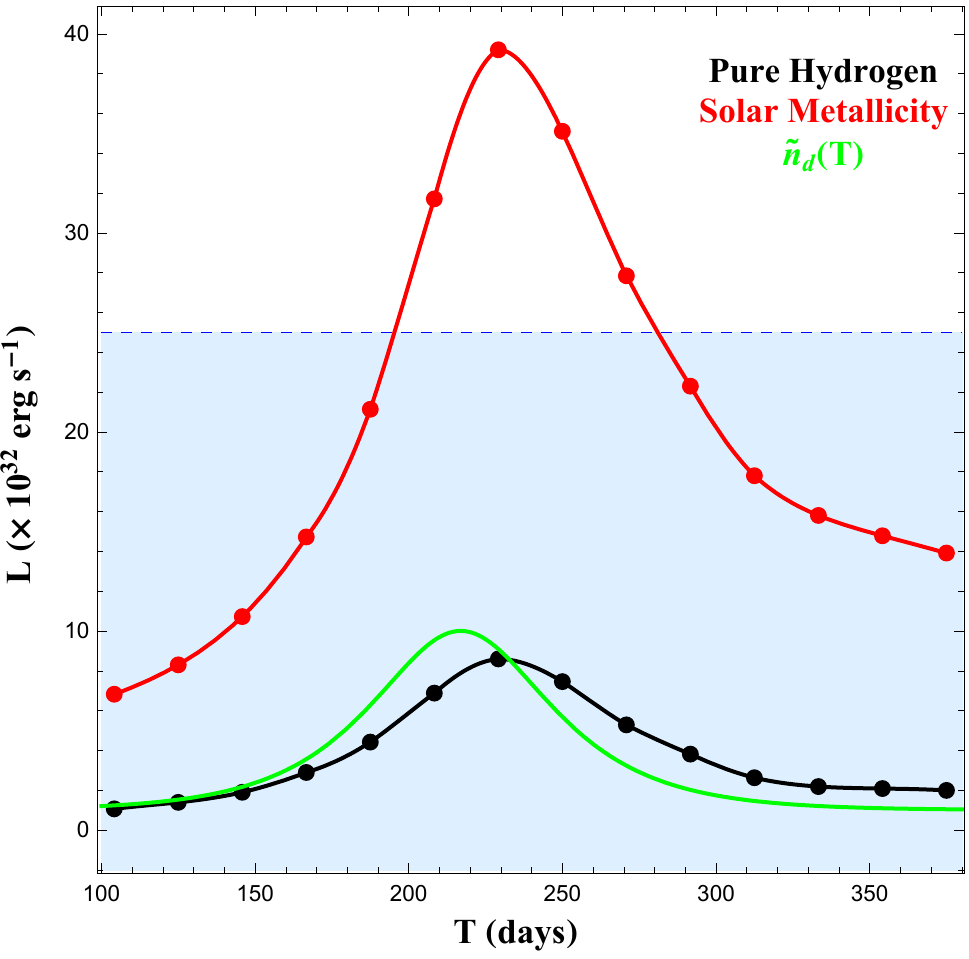}
\caption{Light curves produced from time dependent density simulation for the case $\alpha = 0.1$. The colour coded curves correspond to the different stellar wind compositions as marked on the plot.}
\label{fig:lightcurve}
\end{figure}

Tracking the evolution of the luminosity is completed by calculating the thermal bremsstrahlung emission from several snapshots, as described in Sec.~\ref{sec:power}.
The results are summarized in Fig.~\ref{fig:lightcurve} where the two light curves are plotted for the different wind compositions under consideration. For comparison, the temporal variation of the density is overplotted (green line). The dashed, blue line denotes the X-ray luminosity corresponding to the quiescent emission of Sgr~A*. Any model prediction falling in the blue-shaded region is too low to be distinguished from the quiescent emission.
For the case of solar metallicity, it was found that a month-long flare above the quiescent emission is expected, with a peak reaching $L \approx 4 \times 10^{33} \, {\rm erg} \, {\rm s^{-1}}$. This is a rather conservative value, given that a higher luminosity is expected for hotter disks with $\alpha > 0.1$ (see also Fig.~\ref{fig:cumL}).

Figure~\ref{fig:lightcurve} shows that there is a time-lag between the density and photon temporal profiles. The peak of both light curves  appears at a later time than the peak of the density profile. As stated in Sec.~\ref{sec:khinstabilities}, the back area of mixed fluids that is prone to Kelvin-Helmholtz instabilities, contributes a large portion to the total luminosity of each simulation snapshot. The time-lag between the peak luminosity and the peak density reflects the time required for the tail region to feel the effects of the density changes.

\section{Discussion}
\label{sec:discussion}

This model can be considered an extension of the work performed by GS13. Upon comparing the two models, similarities can be found in the temperature and density of the shocked shell. However, even with these similar values, the two works differ in the amount of bremsstrahlung power produced by the shell. This difference can be explained by the following. The previous work of GS13 assumed that a circular termination shock formed in the wind at a radial distance of $R_0$ and that the shocked fluid radiates on an expansion timescale while traversing a distance of $R_0$ after the termination shock. This is equivalent to having an emitting region with a volume  $V\sim4\pi R_{0}^{3}$. This work has shown that the thickness is instead $H\sim R_0 /4$ meaning that the volume of the emitting region is $V\sim 2\pi R_{0}^{2}H \sim 0.5 \pi R_0^3$. It is an interesting coincidence that when effects of the large back region are included into the total luminosity of the system, the luminosities produced by this work and 
GS13 
are of the same order. 

In Sec.~3.3.3 a scaling relation of the shell's bremsstrahlung power on the model parameters $\dot{M}_{\rm w}, n_{\rm d}$ and $\alpha$ was derived (see eq.~(\ref{scale})). The derivation of a similar scaling relation for the bremsstrahlung power from  the whole emitting volume (i.e., shell plus tail) is not straightforward. The main reason is that the volume of the tail region could not be determined analytically. As long as the tail of mixed fluids can be approximated by a cylindrical volume with cross sectional area $A\propto R_0^2$ and length $\propto R_0$, no large deviations from the scalings in eq.~(\ref{scale}) are expected. This has been verified by running various steady-state simulations for different values of $\dot{M}_{\rm w}$ while keeping all other parameters fixed. It was discovered that the total luminosity scales as $L \propto \dot{M}_{\rm w}^{1.3}$ which verifies our previous statement that the volume of the emitting region roughly scales as $R_{0}^{3}$. By tracking the evolution of the 
total bremsstrahlung power and the disk density during the time-dependent simulation, 
we find $L \propto n_{\rm d}^{0.512}$. The dependence on density is therefore similar to that of eq.~(\ref{scale}).
We note that the scaling relation was determined by performing a linear fit to a plot of $\log(L/L_{33})$ vs. $\log(n_{\rm d}/n_{\rm d, 4})$.

Although no specific model was adopted for the accretion disk during this work, it behooves us to briefly discuss a few of the current proposed models. Through GRMHD simulations, the profile of the number density was shown to scale as $R^{-1}$ with typical values at a radial distance $\sim3000R_{\rm g}$ being $\sim10^{4}$ cm$^{-3}$ \citep{sadowski2013, tchekhovskoy2012, mckinney2012}.  If instead, the accretion flow around Sgr~A* is advection dominated (ADAF), the density profile is steeper, i.e.  $n_{\rm d}\propto R^{-3/2}$. Given that X-ray observations probing the disk density close to the Bondi radius ($R_{\rm b}\simeq 2\times 10^{5}$~$R_{\rm g}$) result in $n_{\rm d, b}=100$ cm$^{-3}$ \citep{baganoff2003}, the density of an ADAF disk at the S2 pericenter distance would be $\sim10^{5}$ cm$^{-3}$ \citep{narayan1994}. This is the peak density reached within our time-dependent simulations. Thus, the results  presented in Sec.~4.1 fall within the latter of the models.

Besides the number density, the thermal pressure of the disk, and therefore $\alpha$, are also model-dependent; however, there are less restrictions on the range of allowed $\alpha$ values. This parameter is not limited to being less than unity and has a substantial effect on the flare's luminosity.  For small $\alpha$ values, such as $\alpha \le 0.1$, a  high density (i.e., $n_{\rm d}\gtrsim 10^5$~cm$^{-3}$)  is required for the  production of an observable X-ray flare due to wind-disk interactions. If the $\alpha$ value becomes larger (i.e. $\alpha > 0.1$), the requirement for a high number density of the disk is relaxed and, in turn, a wider range of density profiles becomes relevant. Interestingly, the thermal pressure in the disk directly affects the thickness and, therefore, the time it takes for the star to cross it. The scale height at the midplane of the disk can therefore be probed by the duration of the flare.

This work, though oriented towards the study of S2's passage through the accretion disk, can be applied to the study of other stars within the S-Cluster. For example, S14, another early type B-star is characterized by a close pericenter occuring next in $2047$ \citep{gillessen2009a}. Doing so might allow one to constrain properties not only of these stars but also the accretion disk at various gravitational radii away from Sgr A*.

Although this work is focused on the thermal emission produced by the shocked wind, it is possible that there are other sources of emission occuring throughout the system. Through the interactions with the stellar wind, the disk undergoes a shock which may accelerate non-thermal particles \citep{sadowski2013}. GS13 explored possible inverse Compton signatures and concluded that detection may be possible in the hard X-ray regime. It has also been discussed that accleration at the termination shock of the wind could produce synchrotron emission with possible detections in the radio and infrared band \citep{ginsburg2016}. However, this is observed for a large stellar mass-loss rate of $\dot{M}_{\rm w} = 10^{-6} M_\odot\, {\rm yr}^{-1}$. This mass-loss rate from S2 is expected to be lower \citep{martins2008}.

Although this work is meant to give a somewhat basic description of S2's pericenter passage, tidal effects and the acceleration of the star during its pericenter transit are ignored. It has been shown, in the case of G2's recent pericenter passage, that tidal stretching dominates the evolution of the shocked fluids close to the pericenter \citep{ballone2013, gillessen2014}. These effects are then relaxed after the pericenter passage has been reached and our approach becomes more accurate. It is expected that as S2 approaches its pericenter, the density of the mixing region increases with respect to our simulated values (see Fig.~2 in \cite{ballone2013}) due to both compression in the cylindrical radial coordinate and stretching in the z coordinate. Tidal forces are, therefore, likely to increase the thermal emission from the shocked region. Moreover, it is interesting to note that if the compression in the cylindrical radial direction is too high, it is possible that the mixing region would decouple from the 
star, as depicted in \cite{ballone2013}.

\section{Summary}
\label{sec:summary}

The formation a bow shock in the stellar winds of a massive star through interactions with a hot dense medium is accurately described through the prescription of momentum-supported bow shocks. The application of this model to the study of S2's pericenter passage as it processes through the accretion disk of Sgr~A* results in the shocked wind cooling via thermal bremsstrahlung emission in the $\sim$keV band. The predicted peak X-ray luminosity is $L \approx 4 \times 10^{33} \, {\rm erg} \, {\rm s^{-1}}$ for a stellar mass-loss rate, disk density, and thermal pressure strength equal to $\dot{M}_{\rm w}= 10^{-7} M_\odot\, {\rm yr}^{-1}$, $n_{\rm d} = 10^{5}$ cm$^{-3}$, and $\alpha=0.1$, respectively. Fits to the numerical results show that the X-ray luminosity scales with the model parameters as $L \propto n_{\rm d}^{0.51} \dot{M}_{\rm w}^{1.3} \alpha^{0.56}$. The estimated value, which may be higher by a factor of 3-4 for hotter disks,  exceeds the flux level of the quiescent X-ray emission from Sgr~A* and is, 
therefore, potentially detectable. The detection (or not) of a month-long $\sim$ keV X-ray flare at the next pericenter passage of S2 (mid-2018) will constrain the density and thermal content of the accretion disk at a few thousand gravitational radii from the supermassive black hole.

\section*{Acknowledgments}
We thank M. A. Aloy for helpful comments. M.P. acknowledges support for this work by NASA through Einstein Postdoctoral
Fellowship grant number PF3~140113 awarded by the Chandra X-ray
Center, which is operated by the Smithsonian Astrophysical Observatory
for NASA under contract NAS8-03060. PM acknowledges the support from the European Research Council (grant CAMAP-259276), and the partial support of grants AYA2013-40979-P and PROMETEO-II-2014-069. We thankfully acknowledge the computer resources, technical expertise and assistance provided by the "Centre de C\`alcul de la Universitat de Val\`encia" through the use of {\emph{Tirant}}, the local node of the Spanish Supercomputation Network.
\appendix
\section[]{Parameter scalings with $\lambda$}
\label{app1}
To derive the scalings  of the density, temperature and thickness of the shocked stellar wind region  with $\lambda$, at the maximum angle where the semi-analytical formalism applies ($\theta=\pi/2$), one must first determine the $\lambda$-dependence of $r\left(\frac{\pi}{2}\right)\equiv r_{\pi/2}$.
It was found by fitting to $\lambda$ that $r_{\pi/2}(\lambda) \approx a - b \lambda$ where $a = 1.726$ and $b = 0.834$.
The density at this angle then depends on $\lambda$ as
\eqb
\rho_{\rm sw, \pi/2} = [(1 - \lambda) \, (a-b\lambda)^{2}]^{-1},
\label{denpi}
\eqe
where $r_{\pi/2}$ was substituted in eq.~(\ref{rho2}).
The temperature of the shocked region is dependent upon $\lambda$ through $v_{\rm w \bot}$, governed by eq.(\ref{v2norm}). At $\pi/2$ we find
\eqb
T_{\rm sw, \pi/2} \propto v^2_{\rm w \bot, \pi/2} \propto \frac{\left[1 - r^{2}_{\pi/2}(\lambda)\right]^2} {\left(\frac{\pi}{2}\right)^{2} (1 - \lambda r_{\pi/2}^{2}(\lambda))^{2} + (1 - r^{2}_{\pi/2}(\lambda))^{2}} \cdot
\eqe
The thickness of the shocked region at $\pi/2$ is written as
\eqb
\frac{H(\pi/2)}{R(\pi/2)} \propto \frac{\left(2 \frac{v_{\rm d}}{v_{\rm w}} + r_{\pi/2}^{2} (1 - \lambda)\right)^2}
{\left[\left(\frac{\pi}{2}\right)^{2} (1 - \lambda r_{\pi/2}^{2}(\lambda))^{2} + (1 - r^{2}_{\pi/2}(\lambda))^{2} \right]^{1/2}}\cdot
\eqe

At an angle of $\theta = 0$, where the initial condition is $r(0) = 1$, the desnity ratio goes as the square of $u\equiv \frac{v_{\rm d}}{v_{\rm w}}$
\eqb
\rho_{\rm sw} / \rho_{\rm d} = 4 u^{2} (1 - \lambda)^{-1}
\eqe 
where we have adopted an adiabatic index of $\gamma = 5/3$. As described in Sec~\ref{ssw}, the temperature of the shocked wind is independent of $\lambda$ at $\theta = 0$
\eqb
T_{\rm sw}(\theta=0) = \frac{3}{32} \frac{k_{\rm b}}{m_{\rm p}} v_{\rm w}^{2}.
\eqe
This comes from the use of the ideal equation of state and the cancelation of the $\lambda$ dependence from both $\rho_{\rm sw}$ and $P_{\rm sw}$.

 \bibliographystyle{mn2e} 
 \bibliography{wind_disk}

\end{document}